\documentclass[12pt]{article}
\begin{document}
\newfont{\elevenmib}{cmmib10 scaled\magstep1}%
\renewcommand{\theequation}{\arabic{section}.\arabic{equation}}
\newcommand{\tabtopsp}[1]{\vbox{\vbox to#1{}\vbox to12pt{}}}

\newcommand{\preprint}{
            \begin{flushleft}
   \elevenmib Yukawa\, Institute\, Kyoto\\
            \end{flushleft}\vspace{-1.3cm}
            \begin{flushright}\normalsize  \sf
            YITP-00-24\\
           {\tt hep-th/0005277} \\ May 2000
            \end{flushright}}
\newcommand{\Title}[1]{{\baselineskip=26pt \begin{center}
            \Large   \bf #1 \\ \ \\ \end{center}}}
\newcommand{\Author}{\begin{center}\large \bf
            S.\, P.\, Khastgir, A.\,J.\, Pocklington and R.\, Sasaki
     \end{center}}
%\hspace*{2.13cm}%
\hspace*{0.7cm}%
\newcommand{\Address}{\begin{center} \it
            Yukawa Institute for Theoretical Physics, Kyoto
            University,\\ Kyoto 606-8502, Japan
      \end{center}}
\newcommand{\Accepted}[1]{\begin{center}{\large \sf #1}\\
            \vspace{1mm}{\small \sf Accepted for Publication}
            \end{center}}
\baselineskip=20pt

\preprint
\thispagestyle{empty}
\bigskip
\bigskip

\Title{Quantum Calogero-Moser Models: \\ Integrability for all Root Systems}
\Author

\Address
\vspace{1.5cm}

\begin{abstract}
The issues related to the integrability of quantum Calogero-Moser models
based on any root systems are addressed.
For the models with degenerate potentials, {\em i.e.}
the rational with/without the harmonic confining force,
the hyperbolic and the trigonometric,
we demonstrate the following for all the root systems:
(i) Construction of a complete set of quantum conserved quantities in terms
of a total sum of the Lax matrix \(L\), {\em i.e.}
\(\sum_{\mu,\nu\in{\cal R}}(L^n)_{\mu\nu}\), in which
\({\cal R}\) is a representation
space of the Coxeter group. (ii) Proof of Liouville integrability.
(iii) Triangularity of the quantum Hamiltonian and the entire
discrete spectrum. Generalised Jack polynomials are defined for all root
systems as unique eigenfunctions of the Hamiltonian. (iv) Equivalence of
the Lax operator and the Dunkl operator. (v) Algebraic construction of all
excited states in terms of creation operators.\  These are mainly
generalisations of the results known for the models based on the
\(A\) series, {\em i.e.} \(su(N)\) type, root systems.
\end{abstract}
\bigskip
\bigskip
\bigskip

%%%%%%%%%%%%%%%
\section{Introduction}
\label{intro}
\setcounter{equation}{0}

Calogero-Moser models are one-dimensional dynamical systems with long
range interactions having a remarkable property that they are integrable
at both classical and quantum levels.
In fact the integrability or more precisely the triangularity of
the quantum Hamiltonian was
first discovered by Calogero \cite{Cal} for the model with inverse square
potential plus a confining harmonic force and by Sutherland \cite{Sut}
for the particles on a circle with the inverse square potential.
Later classical integrability of the models in terms of Lax pairs
was proved by Moser \cite{CalMo}. Olshanetsky and Perelomov \cite{OP1}
showed that these models were based on \(A_r\) root systems and
generalisations of the models based on other root systems including the
non-crystallographic ones were introduced \cite{OP2}.

In this paper we discuss quantum Calogero-Moser models with degenerate
potentials, that is the rational with/without harmonic force, the
hyperbolic and the trigonometric potentials based on all root systems.
We demonstrate, based on previous works of universal Lax pairs for
classical \cite{bcs1,bcs2,DHoker_Phong} and quantum models
\cite{bms}, that various results known for the quantum \(A_r\) models can
be generalised to  the models based on any root systems as well. They are
(i) Construction of a complete set of quantum conserved quantities in
terms of quantum Lax pairs and other methods.
(ii) Universal proof of Liouville integrability for the rational,
hyperbolic and trigonometric potential models.
Namely, the quantum conserved quantities commute among themselves.
(iii) Triangularity of the quantum Hamiltonian
is demonstrated explicitly for all the models.
In other words, the Hamiltonian is shown, in certain bases, to be
decomposed into a sum of finite dimensional triangular matrices.
Thus any eigenvalue equation can be solved by finite steps of linear
algebraic processes only. This also gives the entire discrete spectrum of
the models. As unique eigenfunctions of the Hamiltonian, generalisation of
Jack polynomials and multivariable Laguerre (Hermite) polynomials are
defined for all root systems.
(iv) Equivalence of the quantum Lax pair method and that of so-called
differential-reflection (Dunkl) operators \cite{Dunk} is demonstrated.
(v) For rational models with harmonic confining force, an algebraic
construction of all excited states in terms of creation (annihilation)
operators is achieved.

For the \(A_r\) models, the Lax pairs, conserved quantities and their
involution were discussed by many authors with varied degrees of
completeness and rigour, see for example,
\cite{OP2}, \cite{OP3}--\cite{Ruijs}.
The point (iv) was shown by Wadati and  collaborators \cite{UjWa}
and point (v) was initiated by Perelomov \cite{Pere1} and developed by
Brink and collaborators \cite{Br} and Wadati and  collaborators
\cite{UjWa}. A rather different approach by Heckman and Opdam
\cite{Heck2,HeOp} to Calogero-Moser models with degenerate potentials
based on any root systems should also be mentioned in this connection.

For the general background and the motivations of this
series of papers and the physical applications of the
Calogero-Moser models with various potentials to
lower dimensional physics, ranging from solid state to
particle physics and   supersymmetric Yang-Mills theory, we refer to our
previous papers \cite{bcs1,bcs2} and references therein.

This paper is organised as follows. In section two, quantum Calogero-Moser
Hamiltonian with degenerate potentials is introduced as a
factorised form (\ref{facMHamiltonian}). Connection with root systems and
the Coxeter invariance is emphasised.
Some rudimentary facts of the root systems and reflections are summarised
in Appendix A. A universal Coxeter invariant ground state wavefunction
together with its energy eigenvalue are presented.
In section three we show that all the excited states are also Coxeter
invariant and that the Hamiltonian is {\em triangular} in certain bases.
Complete sets of quantum conserved quantities are
derived from quantum Lax operator \(L\) in section four.
Instead of the trace, the {\em total sum\/} of \(L^n\) is conserved.
That is Ts\((L^n)=\sum_{\mu,\nu\in{\cal R}}(L^n)_{\mu\nu}\), in which
\({\cal R}\) is a representation space of the Coxeter group.
The details of the
complete set for each root system are given in Appendix B.
In section five the creation and annihilation operators for the rational
models with harmonic force are derived.
In section six, the equivalence of the Lax pair operator formalism and
the so-called differential-reflection (Dunkl) operators is demonstrated
and the quantum conserved quantities are expressed in terms of the latter.
In section seven an algebraic construction of excited states
in terms of the differential-reflection (Dunkl) operators for rational
models with harmonic force is presented.
The complete sets of explicit eigenfunctions for the rank two models
are derived in terms of separation of variables based on the Coxeter
invariant variables.
Section eight gives a universal proof of the Liouville integrability for
models with rational (without the confining force), hyperbolic and
trigonometric potentials.
For rational models with harmonic force, the involution is demonstrated
for those based on classical root systems.
A simple use of the quantum Lax pair with spectral parameter is mentioned.
The final section is for summary and comments.

%%%%%%%%%%%%%%%%%%%%%%%%%%%%%%%%%%%%%%%%%%%%%
\section{Quantum Calogero-Moser Models}
\label{cal-mo}
\setcounter{equation}{0}
In this section we briefly introduce the {\em quantum\/} Calogero-Moser
models along with
appropriate notation and background for the main body of this paper.
A   Calogero-Moser model is a
Hamiltonian system associated with a root system \(\Delta\)
of rank \(r\),
which is a set of
vectors in \(\mathbf{R}^{r}\) with its standard inner product.
A brief review of the properties of the root
systems and the associated reflections
will be found in the Appendix A.

\subsection{Factorised Hamiltonian}
The dynamical variables of the Calogero-Moser model are the coordinates
\(\{q_{j}\}\) and their canonically conjugate momenta \(\{p_{j}\}\), with
the canonical commutation relations:
\begin{equation}
   [q_{j},p_{k}]=i\delta_{jk},\qquad [q_{j},q_{k}]=
   [p_{j},p_{k}]=0,\quad j,k=1,\ldots,r.
\end{equation}
These will be denoted by vectors in \(\mathbf{R}^{r}\)
\begin{equation}
   q=(q_{1},\ldots,q_{r}),\qquad p=(p_{1},\ldots,p_{r}).
\end{equation}
The momentum operator \(p_j\) acts as
\[
   p_j=-i{\partial\over{\partial q_j}}, \quad j=1,\ldots,r.
\]
As for the interactions we consider only the degenerate potentials,
that is the rational (with/without
harmonic force), hyperbolic and trigonometric potentials:
\begin{equation}
   V(\rho\cdot q)=
   \left\{\begin{array}{rll}
   {1/{(\rho\cdot q)^2}},& \mbox{type I},&\\
   {a^2/{\sinh^2 a(\rho\cdot q)}}, &\mbox{type II},&\quad \rho\in\Delta,\\
   {a^2/{\sin^2 a(\rho\cdot q)}},&\mbox{type III},&
   \end{array}
   \right.
   \label{potfun}
\end{equation}
in which \(a\) is an arbitrary real positive constant, determining the
period of the trigonometric potentials.
They imply integrability for all of the
Calogero-Moser models based on the crystallographic root systems.
Those models based on the non-crystallographic root systems, the dihedral
group
\(I_2(m)\), \(H_3\), and \(H_4\), are integrable only for  the
rational potential.
The rational potential models are also integrable if a confining
harmonic
potential
\begin{equation}
   {1\over2}\omega^2q^2,\quad \omega>0,\qquad \mbox{type V},
   \label{harmpot}
\end{equation}
is added to the Hamiltonian.
Since we will discuss the universal properties and solutions applicable
to all the interaction types as well as those for specific interaction
potentials, let us adopt the conventional nomenclature for them.
We call the models with rational, hyperbolic, trigonometric and
rational with harmonic force the type I, II, III and V models, respectively.
(Type IV models have elliptic potentials
which we will not discuss in this paper.)

\bigskip
The Hamiltonian for the {\em quantum\/} Calogero-Moser model can be written
in a `factorised form':
\begin{eqnarray}
   \label{facMHamiltonian}
   \mathcal{H} &=& {1\over 2}\sum_{j=1}^r\left(p_j- i\frac{\partial
   W}{\partial q_j}\right)\left(p_j+i{\partial W\over{\partial
   q_j}}\right),\\
   &=&{1\over 2}\sum_{j=1}^{r}\left(p_{j}^{2}+\left({\partial W
   \over{\partial q_{j}}}\right)^{2}\right)
   +{1\over2}\sum_{j=1}^r{\partial^{2}W\over
  {\partial q_{j}^2}},
   \label{HscB}\\
   &=&{1\over 2} p^{2} + {1\over2}\sum_{\rho\in\Delta_+}
   {g_{|\rho|}(g_{|\rho|}-1) |\rho|^{2}}
   \,V(\rho\cdot q)+ ({\omega^2\over2}q^2)-{\cal E}_0.
 \label{qCMHamiltonian}
\end{eqnarray}
It should be noted that the above factorised Hamiltonian
(\ref{qCMHamiltonian})
consists of an operator part \(\hat{\cal H}\),
which is the Hamiltonian in the usual
definition, and a constant
\({\cal E}_0\) which is  the ground state energy to be discussed later:

\begin{eqnarray}
  \mathcal{H}&=&\hat\mathcal{H}-{\cal E}_0,\label{sumHam}\\
  \hat\mathcal{H}&=&{1\over 2} p^{2} + {1\over2}\sum_{\rho\in\Delta_+}
   {g_{|\rho|}(g_{|\rho|}-1) |\rho|^{2}}
   \,V(\rho\cdot q)+ ({\omega^2\over2}q^2).
\label{opHam}
\end{eqnarray}
The real {\em positive\/} coupling constants \(g_{|\rho|}\)
 are defined on orbits of the corresponding
Coxeter group, {\it i.e.} they are
identical for roots in the same orbit. That is, for the simple Lie
algebra cases one coupling constant
\(g_{|\rho|}=g\) for all roots in simply-laced models
and  two independent coupling constants, \(g_{|\rho|}=g_L\)
for long roots and \(g_{|\rho|}=g_S\) for
short roots in non-simply laced models. For the \(I_{2}(m)\) models,
 there is one coupling if \(m\) is odd, and two independent ones if \(m\)
is even. Let us call them \(g_e\) for even roots and \(g_o\) for odd roots.
Throughout this paper we consider the coupling constants at generic values.
We parametrise the positive roots of \(I_2(m)\) as
\begin{equation}
   \rho_j=(\cos((j-1)\pi/m),\sin((j-1)\pi/m)),\quad j=1,\ldots,m.
   \label{dihedroots}
\end{equation}
The \(H_3\) and \(H_4\) models have one coupling constant \(g_{|\rho|}=g\),
since these root systems are simply-laced.
It should be noted that the operator part of the Hamiltonian \(\hat{\cal
H}\)
is strictly positive for \(g_{|\rho|}\ge1\).

The simplest way to introduce the factorised form is through
supersymmetry \cite{bms,FredMend}, in which  function   \(W\) is
called a superpotential:
\begin{equation}
   W(q)=\sum_{\rho\in\Delta_+}g_{|\rho|}\ln|w(\rho\cdot q)|+
   (-{\omega\over2}q^2),
   \quad g_{|\rho|}>0,\quad \omega>0.
   \label{suppot}
\end{equation}
The potential \(V(u)\) (\ref{potfun}) and the function \(w(u)\) are related
by
\begin{eqnarray}
 y(u)&\equiv& {d\over {du}} x(u),\quad
   {dw(u)\over {du}}/w(u)\equiv x(u),\label{hdef}\\
 V(u)&=&-y(u)=x^2(u)+a^2\times\left\{
   \begin{array}{rl}
      0& \mbox{rational},\\
      -1& \mbox{hyperbolic},\\
      1& \mbox{trigonometric}.
   \end{array}\right.
   \label{Vxrel}
\end{eqnarray}
The following Table I gives these functions for each potential:
\begin{center}
  \begin{tabular}{|lc|c|c|c|}
     \hline
      potential& type&\(w(u)\) & \(x(u)\) & \(y(u)\) \\
     \hline
     rational & I \& V&\(u\) & \(1/u\) & -\(1/u^2\) \\
     \hline
     hyperbolic & II&\(\sinh au\) & \(a\coth au\) & -\(a^2/\sinh^2 au\) \\
     \hline
     trigonometric &III& \(\sin au\) & \(a\cot au\) & -\(a^2/\sin^2 au\) \\
     \hline
  \end{tabular}\\
\bigskip
 Table I: Functions appearing in the Lax pair and superpotential.
\end{center}
For proofs that the factorised Hamiltonian (\ref{HscB}) actually gives
the quantum Hamiltonian (\ref{qCMHamiltonian}) for
{\em all the root systems and
potentials\/} see \cite{OP2,OP3,bms}. It is easy to
verify that for any potential
\(V(u)\), the Hamiltonian is invariant under reflection of the phase space
variables in the hyperplane perpendicular to any root
\begin{equation}
  {\cal H}(s_{\alpha}(p),s_{\alpha}(q))={\cal H}(p,q), \quad
   \forall\alpha\in\Delta
  \label{HamCoxinv}
\end{equation}
with \(s_{\alpha}\) defined by (\ref{Root_reflection}).

Some remarks are in order. For all of the root systems and for
any choice of potential (\ref{potfun}),
the  Calogero-Moser model has a hard repulsive potential \(\sim
{1/{(\alpha\cdot q)^2}}\) near the reflection hyperplane
\(H_{\alpha}=\{q\in\mathbf{R}^{r},\, \alpha\cdot q=0\}\).
The strength of the singularity is given by
the coupling constant \(g_{|\alpha|}(g_{|\alpha|}-1)\)
which is {\em independent\/} of the choice of the normalisation of the
roots.
(Thus for rational models with/without harmonic force there is equivalence,
\(A_2\cong I_2(3)\), \(B_2\cong I_2(4)\), \(G_2\cong I_2(6)\),
\(B_r\cong C_r\cong
BC_r\).) This determines the form of the ground state wavefunction, as we
will see in subsection \ref{grfun}.
This repulsive potential is classically insurmountable.
Thus the motion is always
confined within one Weyl chamber.
This  feature allows us to constrain the configuration space to
the principal Weyl chamber (\(\Pi\): set of simple roots, see Appendix A)
\begin{equation}
   PW=\{q\in{\bf R}^r|\ \alpha\cdot q>0,\quad \alpha\in\Pi\},
   \label{PW}
\end{equation}
without loss of generality.
In the case of the trigonometric potential, the configuration space is
further
limited due to the periodicity of the potential to
\begin{equation}
   PW_T=\{q\in{\bf R}^r|\ \alpha\cdot q>0,\quad \alpha\in\Pi,
   \quad \alpha_h\cdot q<\pi/a\},
   \label{PWT}
\end{equation}
where \(\alpha_h\) is the highest root.

The fact that the classical motions are confined in the Weyl
chamber (alcove) \(PW (PW_T)\)  does not necessarily mean that the
corresponding quantum wavefunctions vanish identically outside of
the region.
On the contrary, as we will see soon, the ground state wavefunction
(subsection \ref{grfun}) and all the other excited states
wavefunctions (section \ref{coxinv}) are Coxeter invariant, reflecting
the Coxeter invariance of the Hamiltonian (\ref{HamCoxinv}).
In the early years of Calogero-Moser models in which those based on
the \(A_r\) root system were mainly discussed, these Coxeter invariant
solutions were considered as totally symmetric states of bosonic
systems. We will not, however, adopt this interpretation, for in
the models based on the other root systems the reflection is not the
same as particle interchange.
The quantum theory we are discussing is the so-called first quantised
theory.
That is, the notions of identical particles and the associated statistics
are non-existent.

\subsection{Ground state wavefunction and energy}
\label{grfun}
One merit of the factorised Hamiltonian (\ref{facMHamiltonian}) is
the ease of derivation of the ground state wavefunction
and of the Hamiltonian (\ref{simtraH}) derived by the similarity
transformation in terms of the ground state wavefunction.
Supersymmetric formulation of the Calogero-Moser models
\cite{FredMend,ShasSuth,bms} provides a natural setting for the
introduction of the factorised Hamiltonian. The {\em universal\/} ground
state wavefunction is
\begin{equation}
   \Phi_0(q)=e^{W(q)}=\prod_{\rho\in\Delta_+}
   |w(\rho\cdot q)|^{g_{|\rho|}}\,e^{-{\omega\over2}q^2}.
   \label{grsol}
\end{equation}
The exponential factor \(e^{-{\omega\over2}q^2}\)
exists only for the rational
potential case with the harmonic confining force.
It is easy to see that it is an eigenstate of the Hamiltonian
(\ref{facMHamiltonian}) with {\em zero\/} eigenvalue:
\begin{equation}
   {\cal H}\Phi_0(q)={1\over 2}\sum_{j=1}^r\left(p_j- i\frac{\partial
   W}{\partial q_j}\right)\left(p_j+i{\partial W\over{\partial
   q_j}}\right)\Phi_0(q)=
   0,
\end{equation}
since it satisfies
\begin{equation}
   \left(p_j+i{\partial W\over{\partial
   q_j}}\right)e^{W(q)}=0,\quad  j=1,\ldots,r.
   \label{Weq}
\end{equation}
By using the decomposition of the factorised Hamiltonian into the operator
Hamiltonian (\ref{opHam}) and a constant, we obtain
\begin{equation}
   \hat\mathcal{H}\,e^W\equiv\left({1\over 2} p^{2}
   + {1\over2}\sum_{\rho\in\Delta_+}
   g_{|\rho|}(g_{|\rho|}-1) |\rho|^{2}
   \,V(\rho\cdot q)+({\omega^2\over2}q^2)\right)\,e^W={\cal E}_0\,e^W.
   \label{purbossol}
\end{equation}
In other words, the above solution (\ref{grsol}) provides an eigenstate of
the
Hamiltonian operator \(\hat{\cal H}\) with energy \({\cal E}_0\).
The fact that it is a ground state (for type I, III and V)  can be easily
shown within the framework of the supersymmetric model \cite{bms} thanks to
the
positivity of the supersymmetric Hamiltonian.
It should
be stressed that
\({\cal E}_0\) is determined purely algebraically \cite{bms}, without
really applying the operator on the left hand side of (\ref{purbossol}) to
the
wavefunction.
This type of ground states has been known for some time.
It is derived by various methods, see for example \cite{CalMo,OP2}, and
also by using supersymmetric quantum mechanics
for the models based on classical root systems \cite{FredMend,ShasSuth}.

The ground state energy for the rational potential cases are
\begin{equation}
   {\cal E}_0= \left\{
   \begin{array}{cll}
      0&&\mbox{type I},\\
      \omega\left({r\over2}+\sum_{\rho\in\Delta_+}g_{|\rho|}\right)&&
      \mbox{type V}.
   \end{array}
   \right.
\end{equation}
The same for the hyperbolic  and trigonometric  potential cases are
\begin{equation}
   {\cal E}_0=
   {2a^2}\varrho^2
   \times
   \left\{
   \begin{array}{cl}
      -1&\mbox{hyperbolic},\\
      1&\mbox{trigonometric},
   \end{array}
   \right.
   \label{e0fom}
\end{equation}
in which
\begin{equation}
   \varrho ={1\over2}\sum_{\rho\in\Delta_+}g_{|\rho|}\rho
\end{equation}
can be considered as a `deformed Weyl vector' \cite{OP2,HeOp}.
Again these formulas are universal. That is they apply to all of the
Calogero-Moser models based on any root systems.
A negative \({\cal E}_0\) for the obviously positive Hamiltonian of the
hyperbolic potential model indicates that the interpretation of \(e^W\)
as an eigenfunction is not correct. This function diverges as
\(|q|\to\infty\) for the hyperbolic and the rational
potential cases, destroying the hermiticity of the Hamiltonian.
 Obviously we have
\begin{equation}
   \int_{PW\,(PW_T)}e^{2W(q)}\,dq=\left\{
   \begin{array}{cl}
      \infty&\mbox{: type I and II},\\
      \mbox{finite}&\mbox{: type III and V},
   \end{array}
   \right.
\end{equation}
in which \(PW\) and \(PW_T\) denote that the integration is over the regions
defined in (\ref{PW}) and (\ref{PWT}).
Naturally,  most existing results in quantum Calogero-Moser models are
for the models with trigonometric potential and the
rational potential with harmonic force which have  normalisable states
and discrete spectra.
%There are also some results
%for the rational and hyperbolic models \cite{CalMo,Dunk}.

It should be remarked that the domain of the universal
ground state wavefunction
\(\Phi_0\) could be considered as the  entire \({\bf R}^r\) space
except for the points on the reflection hyperplanes,
that is the disjoint union of all the Weyl chambers (alcoves),
instead of
the initial Weyl  chamber/alcove (\(PW\), \(PW_T\)) in which the
classical motions  are restricted due to the singular potential. In
fact,
\(\Phi_0\) and \(W\) are characterised as {\em Coxeter invariant\/}:
\begin{equation}
   \check{s}_{\rho}\Phi_0=\Phi_0,\quad \check{s}_{\rho}W
   =W,\qquad \forall\rho\in\Delta,
   \label{Coxinv}
\end{equation}
in which   \(\check{s}_{\rho}\) is the representation of the reflection
in the function space. For an arbitrary function \(f\) of \(q\),
its action is defined by
\begin{equation}
   (\check{s}_{\rho}f)(q)=f(s_{\rho}(q)).
   \label{defcheck}
\end{equation}
This definition can be generalised to the entire Coxeter group
\(G_{\Delta}\):
for an arbitrary element \(g\) of \(G_{\Delta}\), \(\check{g}\) is defined
by:
\begin{equation}
   (\check{g}f)(q)=f(g^{-1}(q)),\qquad \forall g\in G_{\Delta}.
\end{equation}
In the rest of this paper we discuss mainly  the type III and V models
 which have  normalisable states
and discrete spectra.

\section{Coxeter invariant excited states, triangularity and spectrum}
\label{coxinv}
\setcounter{equation}{0}
In this section we show that all the excited states wavefunctions
are Coxeter invariant, too.
In other words, the Fock space consists of Coxeter invariant functions
only. With the knowledge of the ground state wavefunction
\(e^W\), the other states of the
Calogero-Moser models can be easily obtained as  eigenfunctions of a
differential operator
\(\tilde{\cal H}\) obtained from
\({\cal
H}\) by a similarity transformation:
\begin{eqnarray}
   \tilde{\cal H}&=&
   e^{-W}\,{\cal H}\,e^W\nonumber\\
   &=&e^{-W}\left({1\over 2} p^{2} +
   {1\over2}\sum_{\rho\in\Delta_+}
   g_{|\rho|}(g_{|\rho|}-1) |\rho|^{2}
   \,V(\rho\cdot q)+({\omega^2\over2}q^2)-{\cal E}_0\right)\,e^W,
\end{eqnarray}
\begin{equation}
   \tilde{\cal H}\Psi_{\lambda}=\lambda\Psi_{\lambda} \quad
   \Longleftrightarrow\quad
   {\cal H}\,\Psi_{\lambda}\,e^W=\lambda\Psi_{\lambda}\,e^W.
   \label{neweq}
\end{equation}
Thanks to the factorised form of the Hamiltonian \({\cal H}\)
(\ref{facMHamiltonian}), (\ref{HscB}), the transformed Hamiltonian
\(\tilde{\cal H}\) takes a simple form:
\begin{equation}
   \tilde{\cal H}=-{1\over2}\sum_{j=1}^r\left(
   {\partial^2\over{\partial q_j^2}}+2{\partial W\over{\partial q_j}}
   {\partial\over{\partial q_j}}\right).
   \label{simtraH}
\end{equation}
The Coxeter invariance of \(W\) implies those of \({\cal H}\) and
\(\tilde{\cal H}\):
\begin{equation}
   \check{s}_{\rho}{\cal H}\check{s}_{\rho}={\cal H}, \quad
   \check{s}_{\rho}\tilde{\cal H}\check{s}_{\rho}=\tilde{\cal H},\qquad
   \forall
   \rho\in\Delta.
   \label{Htilinv}
\end{equation}
For type I and III models we introduce proper bases of Fock space
consisting of Coxeter invariant functions and show that the above
Hamiltonian \(\tilde{\cal H}\) (\ref{simtraH}) is {\em triangular\/} in
these
bases.
This establishes the integrability of the type I and III models
universally
\footnote{Triangularity of the \(A_r\) type V and III Hamiltonians
was noted in the original papers of Calogero \cite{Cal} and Sutherland
\cite{Sut}. That of rank two models in the Coxeter invariant bases was
shown in \cite{HeOp,ruhl}.}
and also gives the entire spectrum  of the Hamiltonian, see
(\ref{omespec}), (\ref{Ndiv}) and (\ref{trigspec}).

\bigskip
\subsection{Rational potential with harmonic force}
\label{harmtri}
First, let us determine the structure of the set of eigenfunctions of the
transformed Hamiltonian \(\tilde{\cal H}\), for the type V models:
\begin{equation}
   \tilde{\cal H}=\omega\, q\cdot\!{\partial\over{\partial q}}
   -{1\over2}\sum_{j=1}^r{\partial^2\over{\partial q_j^2}}
   -\sum_{\rho\in\Delta_+}{g_{|\rho|}\over{\rho\cdot q}}
   \rho\cdot\!{\partial\over{\partial q}}.
   \label{Htilom}
\end{equation}
Obviously a constant and \(\omega q^2-{\cal E}_0/\omega\)
are its eigenfunctions
with eigenvalue 0 and \(2\omega\), respectively.
Let us suppose that a polynomial  \(P(q)\) is an eigenfunction
of \(\tilde{\cal H}\):
\begin{equation}
   \tilde{\cal H}P(q)=\lambda P(q).
   \label{htileigen}
\end{equation}
Due to the Coxeter invariance of \(\tilde{\cal H}\) (\ref{Htilinv}),
we know that \(\check{s}_{\rho}P\) together with the difference
\[
Q=(1-\check{s}_{\rho})P
\]
are also eigenfunctions with the same
eigenvalue:
\begin{equation}
   \tilde{\cal H}Q(q)=\lambda Q(q),
   \label{Qtileq}
\end{equation}
if the latter is not identically zero. Since \(Q\) is a polynomial which is
odd under reflection \(\check{s}_{\rho}\)
\[
   \check{s}_{\rho}Q(q)=-Q(q),
\]
it can be factorised as
\begin{equation}
   Q(q)=(\rho\cdot q)^{2n+1}\tilde{Q}(q),\quad
   \left.\tilde{Q}\right|_{\rho\cdot q=0}\neq0,
   \label{Qtilfac}
\end{equation}
with a non-negative integer \(n\) and a polynomial \(\tilde{Q}\).
By substituting (\ref{Qtilfac}) into  (\ref{Qtileq}) and using the
explicit form of \(\tilde{\cal H}\) near the reflection hyperplane
\(\rho\cdot q=0\), we obtain
\begin{equation}
   -(\rho\cdot q)^{2n-1}(2n+1)(n+g_{|\rho|})\rho^2\tilde{Q}
   +{\cal O}[(\rho\cdot q)^{2n}]=\lambda(\rho\cdot q)^{2n+1}\tilde{Q},
\end{equation}
which would imply the vanishing of \(\tilde{Q}\) on the reflection
hyperplane
\[
    \left.\tilde{Q}\right|_{\rho\cdot q=0}=0,
\]
an obvious contradiction.
Thus we are led to the conclusion that {\em the eigenfunctions are
Coxeter invariant polynomials\/} and that the Hamiltonian ${\cal H}$
(\ref{Htilom}) maps a Coxeter invariant polynomial to another.

An obvious basis in the space of Coxeter invariant polynomials is the
homogeneous polynomials of various degrees.
This basis has a natural order given by the degree.
For a given degree the space
of homogeneous Coxeter invariant polynomials is  finite-dimensional.
The explicit form of
\(\tilde{\cal H}\) (\ref{Htilom}) shows that it is {\em lower triangular\/}
in this basis and the diagonal elements are
\(\omega\times degree\) as given by the first term.
Independent Coxeter invariant polynomials exist
at the degrees \(f_j\) listed in
Table II:
\begin{equation}
   f_j=1+e_j,\quad j=1,\ldots,r,
   \label{coxinvdeg}
\end{equation}
in which \(\{e_j\}\), \(j=1,\ldots,r\), are the {\em exponents\/} of
\(\Delta\). Let us denote them by
\begin{equation}
z_1(q),\ldots,z_r(q);\quad z_j(\kappa q)=\kappa^{f_j}z_j(q).
\label{zbasis}
\end{equation}

\begin{center}
    \begin{tabular}{||c|l||c|l||}
       \hline%\hline
        \(\Delta\)& \(f_j=1+e_j\) &\(\Delta\)& \(f_j=1+e_j\)\\
       \hline
       \(A_r\) & \(2,3,4,\ldots,r+1\) & \(E_8\) & \(2,8,12,14,18,20,24,30\)
\\
       \hline
       \(B_r\) & \(2,4,6,\ldots,2r\) & \(F_4\) & \(2,6,8,12\) \\
       \hline
       \(C_r\) & \(2,4,6,\ldots,2r\) & \(G_2\) & \(2,6\) \\
       \hline
      \(D_r\) & \(2,4,\ldots,2r-2;r\) & \(I_2(m)\) & \(2,m\) \\
      \hline
      \(E_6\) & \(2,5,6,8,9,12\) & \(H_3\) & \(2,6,10\) \\
      \hline
      \(E_7\) & \(2,6,8,10,12,14,18\) & \(H_4\) & \(2,12,20,30\) \\
      \hline
    \end{tabular}\\
 \bigskip
 Table II: The degrees \(f_j\) in which independent Coxeter
invariant polynomials exist.
\end{center}

\bigskip
Thus we arrive at:\\
the quantum Calogero-Moser models with the rational potential and the
harmonic confining force is algebraically solvable for any
(crystallographic and non-crystallographic) root system \(\Delta\).
The spectrum of the operator Hamiltonian \(\hat{\cal H}\) is
\begin{equation}
   \omega N+{\cal E}_0,
   \label{omespec}
\end{equation}
with a non-negative integer \(N\) which can be expressed as
\begin{equation}
   N=\sum_{j=1}^rn_j f_j,\quad n_j\in{\bf Z}_+,
   \label{Ndiv}
\end{equation}
and
the degeneracy of the above eigenvalue (\ref{omespec}) is the number
of different solutions of (\ref{Ndiv}) for given \(N\).
This is generalisation of Calogero's original argument
for the \(A_r\) model \cite{Cal} to
the models based on arbitrary root systems.
Now let us denote by $\vec{N}$ the set of non-negative integers in
(\ref{Ndiv}):
\begin{equation}
   \vec{N}=(n_1,n_2,\ldots,n_r),
   \label{Nvec}
\end{equation}
and by $\phi_{\vec{N}}(q)$ the homogeneous Coxeter invariant polynomial
of determined by $\vec{N}$ and the above basis $\{z_j\}$ (\ref{zbasis}):
\begin{equation}
\phi_{\vec{N}}(q)=\prod_{j=1}^rz_j^{n_j}(q).
\end{equation}
As shown above, there exists a unique eigenstate $\psi_{\vec{N}}(q)$ for
each $\phi_{\vec{N}}(q)$:
\begin{eqnarray}
\psi_{\vec{N}}(q)&=&\phi_{\vec{N}}(q)+\sum_{\vec{N}^\prime<\vec{N}}
d_{\vec{N}^\prime} \phi_{\vec{N}^\prime}(q),\quad d_{\vec{N}^\prime}:\
const,\\
 \tilde{\cal H}\psi_{\vec{N}}(q)&=&\omega N\psi_{\vec{N}}(q).
\end{eqnarray}
It satisfies the orthogonality relation
\begin{equation}
(\psi_{\vec{N}},\phi_{\vec{N}^\prime})=0,\quad \vec{N}^\prime<\vec{N},
\end{equation}
with respect to the inner product in $PW$:
\begin{equation}
(\psi,\varphi)=\int_{PW}\psi^*(q)\varphi(q)\,e^{2W(q)}\,dq.
\end{equation}
These polynomials $\{\psi_{\vec{N}}(q)\}$ are generalisations of the
multivariable Laguerre (Hermite) polynomials \cite{Lass} known for the
$A_r$ ($B_r$, $D_r$) root systems to arbitrary root systems.

\bigskip
Some remarks are in order.
\begin{enumerate}
\item
  There is no Coxeter invariant linear
  function in \(q\).
  The quadratic invariant polynomial \(q^2=q\cdot q\) exists in all the
  root systems. This corresponds to the universal fact that
  \(f_1=2, (e_1=1)\) for all the root systems.
  Moreover, this is related to the fact that a special sub-series of
  the excited states with \(N=2n_1=n_1f_1\), \(n_j=0,\) \(j\ge2\), can be
  expressed universally in terms of Laguerre polynomials in \(q^2\).
  This will be discussed at the end of section \ref{alconst1} and in
  subsection
  \ref{B2plus}.
\item
  The other Coxeter invariants corresponding to the degrees
\(f_2,\ldots,f_r\)
  could be interpreted as special  `angular' variables of a unit sphere
  \(S^{r-1}\) (\(q^2=1\)), with the first Coxeter
invariant \(\sqrt{q^2}\) being the
  radial coordinate.
  These would provide proper variables for describing solutions.
  Solutions in terms of
  separation of variables are in general possible only for the simplest
  cases, that is the rank two models, which will be demonstrated in
  subsection \ref{ranktwo}.
\item
  For \(\Delta=A_1\), the simplest root system of rank one, the Hamiltonian
  \(\tilde{\cal H}\) can be rewritten in terms of a Coxeter invariant
variable
  \(u=\omega q^2\) as:
  \begin{equation}
     \tilde{\cal H}=\omega
     q{d\over{dq}}-{1\over2}{d^2\over{dq^2}}-{g\over{q}}{d\over{dq}}
=-2\omega\left\{u{d^2\over{du^2}}+(g+{1\over2}-u){d\over{du}}\right\}.
     \end{equation}
     The Laguerre polynomial satisfying the differential equation
     \begin{equation}
     \left\{u{d^2\over{du^2}}+(g+{1\over2}-u){d\over{du}}+n\right\}
    L_n^{(g-{1\over2})}(u)=0,
     \label{lagueeq}
  \end{equation}
  provides an eigenfunction with eigenvalue \(2\omega n\), which corresponds
  to the eigenvalue \(2\omega n+{\cal E}_0\) of \(\hat{\cal H}\).
  This is a well-known result.
\item
  Triangularity of the type I models is also obvious from the above
  argument.
\end{enumerate}

\subsection{Trigonometric potential}
\label{triginvcase}
Here we consider those root systems associated with Lie algebras.
In order to determine the excited states of the type III models,
 we have to consider the periodicity.
The superpotential \(W\) and the Hamiltonian \({\cal H}\) are invariant
under
the following translation:
\begin{equation}
   W(q^\prime)=W(q),\quad
   {\cal H}(p, q^\prime)={\cal H}(p, q),\quad
   q^\prime=q+l^\vee\pi/a,
\end{equation}
in which \(l^\vee\) is an element of the dual weight lattice, that is
\begin{equation}
   l^\vee=\sum_{j=1}^r l_j{2\over{\alpha_j^2}}\lambda_j,\quad l_j\in
   {\bf Z},\quad \alpha_j\in\Pi,\quad
   \alpha_j^\vee\!\cdot\lambda_k=\delta_{jk}.
\end{equation}
As is well known in quantum mechanics with periodic potentials,
the wavefunctions diagonalising the translation operators
are expressed as
\begin{equation}
   e^{2ia\mu\cdot q}\left(\sum_{\alpha\in
   L(\Delta)}b_{\alpha}e^{2ia\alpha\cdot q}\right)e^W,\quad b_\alpha :
   const,
   \quad L(\Delta): \mbox{root lattice},
\end{equation}
in which a vector \(\mu\in{\bf R}^r\) is as yet unspecified. In
other words, up to the overall phase factor \(e^{2ia\mu\cdot q}\), this
is a Fourier expansion in terms of the simple roots.

Let \(P_T(q)\) be a polynomial in \(e^{\pm2ia\alpha_j\cdot q}\),
\(\alpha_j\in \Pi\) and   suppose that a function  \(\phi(q)\)
\begin{equation}
   \phi(q)=e^{2ia\mu\cdot q}P_T(q),\quad \mu\in{\bf R}^r,
   \label{trigeigen}
\end{equation}
is an eigenfunction
of \(\tilde{\cal H}\):
\begin{equation}
   \tilde{\cal H}\phi(q)=\lambda \phi(q),
   \end{equation}
   in which the explicit form of \(\tilde{\cal H}\) is given by
   \begin{equation}
   \tilde{\cal H}=
   -{1\over2}\sum_{j=1}^r{\partial^2\over{\partial q_j^2}}
   -a\sum_{\rho\in\Delta_+}{g_{|\rho|}\cot{(a\rho\cdot q)}}
   \rho\cdot\!{\partial\over{\partial q}}.
   \label{Htilcot}
\end{equation}

As above, due to the Coxeter invariance of \(\tilde{\cal H}\)
(\ref{Htilinv}),  we know that \(\check{s}_{\rho}\phi\) together with
the difference
\[
   \varphi=(1-\check{s}_{\rho})\phi
\]
are also eigenfunctions with the same
eigenvalue:
\begin{equation}
   \tilde{\cal H}\varphi(q)=\lambda \varphi(q).
   \label{phitileq}
\end{equation}
Since \(\varphi\)  is
odd under reflection \(\check{s}_{\rho}\)
\[
   \check{s}_{\rho}\varphi(q)=-\varphi(q),
\]
it can be expressed as
\begin{equation}
   \varphi(q)=(\rho\cdot q)^{2n+1}\tilde{\varphi}(q)+{\cal O}[(\rho\cdot
   q)^{2n+3}],\quad
   \left.\tilde{\varphi}\right|_{\rho\cdot q=0}\neq0,
   \label{phitilfac}
\end{equation}
in a neighbourhood of the reflection hyperplane \(\rho\cdot q=0\).
In this neighbourhood, the singularity structure of
\(\tilde{\cal H}\) for the trigonometric potential is the same as
that of \(\tilde{\cal H}\) for the rational potential discussed in
the previous subsection. Thus we obtain, as before, a contradiction
\(\left.\tilde{\varphi}\right|_{\rho\cdot q=0}=0\).
In other words, the eigenfunction \(\phi\) (\ref{trigeigen}) must
be {\em Coxeter invariant\/}.
This in turn requires that the unspecified vector \(\mu\) in
(\ref{trigeigen}) to be an element of the weight lattice
\begin{equation}
   \mu\in\Lambda(\Delta).
   \label{muinwei}
\end{equation}
Since
\[
   \check{s}_{\rho}\phi(q)=e^{2ia\, s_{\rho}(\mu)\cdot
   q}\check{s}_{\rho}P_T(q)=
   e^{2ia (\mu\cdot q-\rho^\vee\!\cdot\mu \rho\cdot
   q)}\check{s}_{\rho}P_T(q),
\]
the following condition is necessary, but not sufficient,
\begin{equation}
   \rho^\vee\!\cdot\mu\in{\bf Z},\quad \forall \rho\in\Delta
\end{equation}
for the Coxeter invariance of \(\phi\). Thus we arrive at
(\ref{muinwei}).

Let us introduce a basis for the Coxeter invariant
functions of the form (\ref{trigeigen}). Let \(\lambda\) be a
dominant weight
\begin{equation}
   \lambda=\sum_{j=1}^rm_j\lambda_j,\quad m_j\in{\bf Z}_+,
\end{equation}
and \(W_{\lambda}\) be the orbit of \(\lambda\)
by the action of the Weyl group:
\begin{equation}
   W_{\lambda}=\{\mu\in\Lambda(\Delta)|\quad \mu=g(\lambda),\quad
   \forall g\in G_{\Delta}\}.
\end{equation}
We define
\begin{equation}
   \phi_{\lambda}(q)\equiv\sum_{\mu\in W_{\lambda}}e^{2ia\mu\cdot q},
\end{equation}
which is Coxeter invariant. The set of functions
\(\{\phi_{\lambda}\}\)  has an order
\(>\):
\begin{equation}
   |\lambda|^2>|\lambda^\prime|^2\quad
   \Rightarrow\quad \phi_{\lambda}>\phi_{\lambda^\prime}.
\end{equation}
Next we show that \(\tilde{\cal H}\) is {\em lower triangular\/}
in this basis. By using (\ref{Htilcot}) we obtain
\begin{equation}
   \tilde{\cal H}\phi_{\lambda}=2a^2\lambda^2\phi_{\lambda}
   -2ia^2\sum_{\rho\in\Delta_+}\sum_{\mu\in
   W_{\lambda}}{g_{|\rho|}\cot{(a\rho\cdot q)}}
   (\rho\cdot\mu) e^{2ia\mu\cdot q}.
   \label{Htillam}
\end{equation}
First let us fix one positive root \(\rho\) and a weight \(\mu\)
in \(W_{\lambda}\)
such that \(\rho\cdot\mu\neq0\).
Then
\begin{equation}
   \mu^\prime\equiv s_{\rho}(\mu)=\mu-(\rho^\vee\!\cdot\mu)\rho\in
W_{\lambda},
   \quad \rho\cdot\mu^\prime=-\rho\cdot\mu.
\end{equation}
Without loss of generality we may assume
\begin{equation}
   \rho^\vee\!\cdot\mu=k>0,\quad k\in{\bf Z}.
\end{equation}
The contribution of the pair \((\mu,\mu^\prime)\) in the summation of
(\ref{Htillam}) reads
\begin{eqnarray}
   &&|\rho\cdot\mu|e^{2ai\mu\cdot q}(1-e^{-2aik\rho\cdot
   q})\cot(a\rho\cdot q)\nonumber\\
   &=&i|\rho\cdot\mu|\left(e^{2ai\mu\cdot q}+e^{2ai\mu^\prime\cdot q}
   +2\sum_{j=1}^{k-1}e^{2ai(\mu-j\rho)\cdot q}\right),
\end{eqnarray}
which is the generalisation of Sutherland's fundamental
identity eq(15) in \cite{Sut}
to arbitrary root systems.
The summation in the expression correspond to
\(\phi_{\lambda^\prime}\) with \(\lambda^\prime\) being lower than
\(\lambda\).
Thus (\ref{Htillam}) reads
\begin{equation}
   \tilde{\cal H}\phi_{\lambda}=2a^2\lambda^2\phi_{\lambda}
   +2a^2\sum_{\rho\in\Delta_+}\sum_{\mu\in
   W_{\lambda}}{g_{|\rho|}}
   |\rho\cdot\mu| e^{2ia\mu\cdot q}
   +\sum_{|\lambda^\prime|<|\lambda|}
   c_{\lambda^\prime}\phi_{\lambda^\prime},
   \label{Htillam2}
\end{equation}
in which \(\{c_{\lambda^\prime}\}\)'s are constants.
It is easy to see that (\(\mu=g(\lambda)\), \(\exists g\in G_{\Delta}\))
\begin{equation}
   \sum_{\rho\in\Delta_+}{g_{|\rho|}}
   |\rho\cdot\mu| =\sum_{\rho\in\Delta_+}{g_{|\rho|}}
   |g(\rho)\cdot\lambda|=(\sum_{\rho\in\Delta_+}{g_{|\rho|}}
   \rho)\cdot\lambda=2\varrho\cdot\lambda,
\end{equation}
which is independent of \(\mu\).
Thus we have demonstrated the triangularity of \(\tilde{\cal H}\):
\begin{equation}
   \tilde{\cal
   H}\phi_{\lambda}=2a^2(\lambda^2+2\varrho\cdot\lambda)\phi_{\lambda}
   +\sum_{|\lambda^\prime|<|\lambda|}c_{\lambda^\prime}
   \phi_{\lambda^\prime},
   \label{Htillam3}
\end{equation}
or that of \(\hat{\cal H}\)
\begin{equation}
   \hat{\cal H}\phi_{\lambda}e^W=2a^2(\lambda+\varrho)^2\phi_{\lambda}e^W
   +\sum_{|\lambda^\prime|<|\lambda|}c_{\lambda^\prime}
   \phi_{\lambda^\prime}e^W,
   \label{Hhatlam3}
\end{equation}
with the eigenvalue
\begin{equation}
   2a^2(\lambda+\varrho)^2.
   \label{trigspec}
\end{equation}
In other words, for each dominant weight \(\lambda\) there exists an
eigenstate of \(\tilde{\cal H}\) with eigenvalue proportional to
$\lambda(\lambda+2\varrho)$. Let us denote this eigenfunction
by $\psi_{\lambda}(q)$:
\begin{eqnarray}
\psi_{\lambda}(q)&=&\phi_{\lambda}(q)+\sum_{|\lambda|^\prime<|\lambda|}
d_{\lambda^\prime} \phi_{\lambda^\prime}(q),\quad d_{\lambda^\prime}:\
const,\\
 \tilde{\cal
H}\psi_{\lambda}(q)&=&2a^2\lambda(\lambda+2\varrho)\psi_{\lambda}(q),
\end{eqnarray}
and call it a {\em generalised Jack polynomial\/}
\cite{Stan}-\cite{Awat}. It satisfies the orthogonality relation
\begin{equation}
(\psi_{\lambda},\phi_{\lambda^\prime})=0,\quad |\lambda|^\prime<
|\lambda|,
\end{equation}
with respect to the inner product in $PW_T$:
\begin{equation}
(\psi,\varphi)=\int_{PW_T}\psi^*(q)\varphi(q)\,e^{2W(q)}\,dq.
\end{equation}
In the \(A_r\) model, specifying a dominant weight $\lambda$ is the same
as giving a Young diagram which designates a Jack polynomial.
It should be emphasised, however, that $\{\psi_{\lambda}\}$ are not
identical to the Jack polynomials even for the $A_r$ root systems,
because of  different
treatments of the center of mass coordinates.
 Detailed properties of these polynomials for
various root systems will be
published elsewhere.

\bigskip
Thus we arrive at:\\
the quantum Calogero-Moser models with the trigonometric potential  are
algebraically solvable for any crystallographic
root system \(\Delta\). The spectrum of the Hamiltonian \(\hat{\cal H}\)
is given by (\ref{trigspec})
in which \(\lambda\) is an arbitrary dominant weight.
This is generalisation of Sutherland's original argument \cite{Sut} to
the models based on arbitrary root systems.

\bigskip
Some remarks are in order.
\begin{enumerate}
\item  The weights \(\mu\) appearing in the lower order terms
\(\{\phi_{\lambda^\prime}\}\)'s are those weights contained in the
Lie algebra representation belonging to the highest weight
\(\lambda\).
\item
As a simple corollary we find that for a minimal weight
\(\lambda\),
\[
   \psi_{\lambda}(q)=\phi_{\lambda}(q)=\sum_{\mu\in
   W_{\lambda}}e^{2ia\mu\cdot q}
\]
 is an eigenfunction of \(\tilde{\cal H}\).
A minimal representation \cite{bcs1} consists of a single Weyl orbit
and all of
its weights \(\mu\) satisfy \(\rho^\vee\!\cdot\mu=0,\pm1\), \(\forall
\rho\in\Delta\).
\item
If \(\lambda=\alpha_h\), the highest root of a simply-laced root
system, \(W_\lambda\) is the set of roots itself. Then the lower
order terms are constants only. We find that
\[
  \psi_{\alpha_h}(q)= 2\sum_{\rho\in\Delta_+}\!
   \left(\cos{(2a\rho\cdot
   q)}+g\rho^2/\alpha_h\cdot(\alpha_h+2\varrho)\right)
\]
is an eigenstate of \(\tilde{\cal H}\).
\item
If \(\lambda=\alpha_{Sh}\), the highest short root of a non-simply
laced root system, \(W_\lambda\) is the set of short roots itself.
The lower
order terms are constants, too.
Similarly as above, we find that
\[
  \psi_{\alpha_{Sh}}(q)= 2\sum_{\rho\in{\Delta_L}_+}\!
   \cos{(2a\rho\cdot q)}+2\sum_{\rho\in{\Delta_S}_+}\!
   \left(\cos{(2a\rho\cdot
   q)}+g_S\rho_S^2/\alpha_{Sh}\cdot(\alpha_{Sh}+2\varrho)\right)
\]
is an eigenstate of \(\tilde{\cal H}\). Here \(\Delta_{L(S)}\) is the
set of long (short) roots.
\item
If \(-\lambda\notin W_{\lambda}\) then
there is   another set of functions containing the weight
\(-\lambda\) which belongs to  the same eigenvalue.
\item
The  Coxeter invariant trigonometric polynomials specified by
the fundamental weights \(\{\lambda_j\}\)
\begin{equation}
   \phi_{\lambda_j}(q)=\sum_{\mu\in
   W_{\lambda_j}}e^{2ia\mu\cdot q},\quad \lambda_j:\
\mbox{fundamental weight},
   \quad j=1,\ldots,r
\end{equation}
are expected to play the role of the fundamental
variables \cite{HeOp,ruhl}.
\item
Let us consider the well-known case \(\Delta=A_1\),
the simplest  root system of
rank one. By rewriting the Hamiltonian \(\tilde{\cal H}\)
in terms of the Coxeter
invariant variable \(z=\cos(a\rho q)\), we obtain
\begin{equation}
   \hat{\cal H}=-{1\over2}{d^2\over{dq^2}}-ag\rho\cot(a\rho
    q){d\over{dq}}=
   -{1\over2}a^2|\rho|^2\left\{(1-z^2){d^2\over{dz^2}}-
    (1+2g)z{d\over{dz}}\right\}.
\end{equation}
The Gegenbauer polynomials \cite{OP2}, a special
case of Jacobi polynomials \(P_n^{(\alpha,\beta)}\) provide
eigenfunctions:
\begin{equation}
   P_n^{(g-{1\over2}, g-{1\over2})}\left(\cos(a\rho q)\right),\quad
   {\cal E}=a^2|\rho|^2(n+g)^2/2,\quad n\in{\bf Z}_+.
\end{equation}
The Jacobi polynomial \(P_{\ell}^{(\alpha,\beta)}(z)\)
satisfies differential
equation
\begin{equation}
   \left\{(1-z^2){d^2\over{dz^2}}+\beta-\alpha
   -(2+\alpha+\beta)z{d\over{dz}}
   +\ell(\ell+\alpha+\beta+1)\right\}P_{\ell}^{(\alpha,\beta)}(z)=0.
   \label{jacobieq}
\end{equation}
Here we follow the notation of \cite{Erd}. They form orthogonal
polynomials with weight \(e^{2W}=|\sin(a\rho q)|^{2g}\) in the
interval \(q\in[0,\pi/a\rho]\), (\ref{PWT}).
The Gegenbauer polynomials have a definite parity, \((-1)^n\),
reflecting the periodicity.
The fundamental period is \(\pi/a\rho\), the length of the interval itself.
The (odd) even degree ones corresponding to (half-odd) integer
spin representations
are (anti-) periodic.
\item
Triangularity of type II models follows from the same algebraic reasoning.
\end{enumerate}

%%%%%%%%%%%%%%%%%%%%%%%%%%%%%%%%%%
\section{Quantum Lax pair and quantum conserved
quantities}
\label{cons}
\setcounter{equation}{0}
Historically, Lax pairs for Calogero-Moser models were presented in terms
of Lie algebra representations \cite{CalMo,OP2}, in particular, the
vector representation of the \(A_r\) models.
However, the invariance of Calogero-Moser models is that of Coxeter group
but not that of the associated Lie algebra, which does not exist for the
non-crystallographic root systems.
Thus the universal and Coxeter covariant Lax pairs are given in terms of
the representations of the Coxeter group.

\subsection{General case}
\label{genconv}
Here we recapitulate the essence of the quantum
Lax pair operators for the Calogero-Moser models with
degenerate potentials and without spectral parameter.
The case with spectral parameter will be discussed briefly in
subsection \ref{laxxi} in connection with the proof of involution of
the quantum conserved quantities for \(A_r\) model \cite{Pol}.
 The quantum Lax pair in this subsection applies
to all the degenerate potential
cases except for the case of the rational potential with the harmonic force,
which will be treated separately in subsection \ref{harmL}.
For details and a full exposition, see \cite{bms}.
The Lax operators without spectral parameter  are
\begin{eqnarray}
   L(p,q) &=& p\cdot\hat{H}+X(q),\qquad X(q)=
   i\sum_{\rho\in\Delta_{+}}g_{|\rho|}
   \,(\rho\cdot\hat{H})x(\rho\cdot
q)\hat{s}_{\rho},
   \label{LaxOpDef}\\
   M(q) &=&
   {i\over2}\sum_{\rho\in\Delta_{+}}g_{|\rho|}|\rho|^2\,y
   (\rho\cdot
q)\,\hat{s}_{\rho}-{i\over2}\sum_{\rho\in\Delta_{+}}g_{|\rho|}|\rho|^2\,y
   (\rho\cdot q)\times I,
   \label{Mtildef}
\end{eqnarray}
in which \(I\) is the identity
operator and \(\{\hat{s}_{\alpha},\,\alpha\in\Delta\}\) are the reflection
operators  of the root system.
In contrast with $\{\check{s}_{\alpha}\}$ operators
(\ref{defcheck}) which act in function space,
\(\{\hat{s}_{\alpha}\}\) act on a set of
\(\mathbf{R}^{r}\) vectors
\({\cal R}=\{\mu^{(k)}\in\mathbf{R}^{r},\ k=1,\ldots, d\}\), permuting them
under the action of the reflection group.
The vectors in \(\cal R\) form a basis for
the representation space \(\bf V\) of dimension \(d\).
%It should be remarked that although \(X\) depends on the spectre parameter
%\(\xi\), the operator \(M\) does not. It also satisfies
The operator \(M\) satisfies the relation
\begin{equation}
   \sum_{\mu\in{\cal R}}M_{\mu\nu}=
   \sum_{\nu\in{\cal R}}M_{\mu\nu}=0,
   \label{sumMzero}
\end{equation}
which is essential for deriving quantum conserved quantities.
The matrix elements of the operators
\(\{\hat{s}_{\alpha},\,\alpha\in\Delta\}\) and \(\{\hat{H}_{j},\,j=1,\ldots,
r\}\) are defined as follows:
\begin{equation}
   (\hat{s}_{\rho})_{\mu\nu}=\delta_{\mu,s_\rho(\nu)}=
      \delta_{\nu,s_\rho(\mu)}, \quad
   (\hat{H}_{j})_{\mu\nu}=\mu_j\delta_{\mu\nu},\quad \rho\in\Delta,
   \quad
   \mu, \nu\in{\cal R}.
   \label{sHdef}
\end{equation}
The form of the function \(x\) depends on the chosen potential, and
the function \(y\)  are defined by (\ref{hdef}), (\ref{Vxrel}).
Note that these relations are only valid for the degenerate potentials
(\ref{potfun}).

The underlying idea of the Lax operator \(L\), (\ref{LaxOpDef}),
is quite simple.
As seen from (\ref{L2Ham}), \(L\) is a ``square root"
of the Hamiltonian.
Thus one part of \(L\) contains \(p\) which is not associated with
roots and another part contains \(x(\rho\cdot q)\), a ``square root"
of the potential
\(V(\rho\cdot q)\), which being associated with a root \(\rho\)
is therefore accompanied by the reflection operator \(\hat{s}_{\rho}\).
Another explanation is the factorised Hamiltonian \({\cal H}\)
(\ref{facMHamiltonian}). We obtain, roughly speaking, \(L\sim\sqrt{\cal
H}\sim p+i{\partial W\over{\partial q}}\hat{s}\) and the property of
reflection
\(\hat{s}^2=1\) explains the sign change in the first term in
(\ref{facMHamiltonian}).

It is straightforward to show that the quantum Lax equation
\begin{equation}
   \label{LaxEquation}
   {d\over dt}{L}=i[{\cal H},L]=[L,{M}],
\end{equation}
is equivalent to the quantum equations of motion derived from the
Hamiltonian (\ref{qCMHamiltonian}). From this it follows:
\begin{equation}
   {d\over{dt}}(L^n)_{\mu\nu}=i[{\cal H},(L^n)_{\mu\nu}]
   =[L^n,M]_{\mu\nu}
   \nonumber\\
   =\sum_{\lambda\in{\cal R}}\left(\phantom{\mbox{\huge
   H}}\hspace{-15pt}
   (L^n)_{\mu\lambda}M_{\lambda\nu}-
   M_{\mu\lambda}(L^n)_{\lambda\nu}\right),\quad
   n=1,\ldots .
\label{qlaxL}
\end{equation}
Thanks to the property of the \(M\) operator
(\ref{sumMzero}):
\[
   \sum_{\mu\in{\cal R}}M_{\mu\nu}=
   \sum_{\nu\in{\cal R}}M_{\mu\nu}=0,
\]
we obtain {\em quantum conserved quantities\/} as the {\em total sum\/}
\((\mbox{Ts})\) of all the
matrix elements of \(L^n\)
\footnote{This type of conserved quantities is known for \(A_r\) models
\cite{ShasSuth,UjWa}.}:
\begin{equation}
   Q_n=\mbox{Ts}(L^n)\equiv\sum_{\mu,\nu\in{\cal R}}(L^n)_{\mu\nu},
  \qquad [{\cal H},Q_n]=0,\quad
   n=1,\ldots.
   \label{qconv}
\end{equation}
Independent conserved quantities appear at such power  \(n\) that
\begin{equation}
   n=1+exponent
\end{equation}
of each root system.
 These are the degrees at which independent Coxeter
invariant polynomials exist.
There are \(r\) exponents for each root system
\(\Delta\) of rank \(r\). Thus we have \(r\)
independent conserved quantities
in Calogero-Moser models. We list in Table II these powers for each root
system.
In particular, the power 2 is universal to all the root systems and
the quantum Hamiltonian (\ref{qCMHamiltonian}) is given by
\begin{equation}
   {\cal H} = {1\over {2 C_{\cal R}}}\mbox{Ts}(L^{2}) +
   const,
   \label{L2Ham}
\end{equation}
where the constant \(C_{\cal R}\) is the quadratic Casimir invariant, which
depends on the
 representation. It is
defined by
\begin{equation}
   \mbox{Tr}(\hat{H}_{j}\hat{H}_{k})
   \equiv\sum_{\mu\in{\cal R}}(\hat{H}_{j}\hat{H}_{k})_{\mu\mu}
   =\sum_{\mu\in{\cal R}}\mu_{j}\mu_{k}
   =C_{\cal R}\,\delta_{jk}.
\end{equation}

\bigskip
Some remarks are in order.
\begin{enumerate}
\item
The Lax pair is Coxeter covariant:
\begin{equation}
   L\left(s_{\rho}(p),s_{\rho}(q)\right)_{\mu\nu}=
   L\left(p,q\right)_{\mu^\prime\nu^\prime},\
   M\left(s_{\rho}(q)\right)_{\mu\nu}=
   M(q)_{\mu^\prime\nu^\prime},\quad \mu^\prime\equiv s_{\rho}(\mu),
   \  \nu^\prime\equiv s_{\rho}(\nu),
\end{equation}
which ensures the Coxeter invariance of the conserved quantities.
\item
Lax pairs can be written down in various representations and the
quantum conserved quantities \(Q_n\) do depend on the representations,
in general.
If necessary, we denote by \(Q_n^{\cal R}\)
the explicit representation dependence.
\item
The availability of plural representations of the Lax pair and the conserved
quantities is essential
for the completeness of the set of conserved quantities as polynomials of
the momentum operators.
For example, let us consider the case of \(D_r\) with even \(r\),
which has two
independent conserved quantities at power \(r\), see Table II.
At least two different representations of the Lax pair are necessary in
order to represent them in the form of (\ref{qconv}).
Those based on the vector, and the (anti)-spinor representations
give two independent conserved quantities. For  \(D_4\)
case, we obtain
\begin{equation}
   Q_4^v=2\sum_{j=1}^4p_j^4,\quad Q_4^s-Q_4^a=24\prod_{j=1}^4p_j,
\end{equation}
in which \(v\), \(s\) and \(a\) stand for the vector, spinor and anti-spinor
representations and we have set \(g=0\) for simplicity.
If two conserved quantities are independent for zero coupling constants,
surely they are so at non-vanishing couplings. Here we have used an explicit
parametrisation of the \(D_r\) root system:
\begin{equation}
   D_r\ \mbox{root system :}\quad \Delta
   =\{\pm e_j\pm e_k,\quad j,k=1,\ldots,r|
   e_j\in{\bf R}^r, e_j\cdot e_k=\delta_{jk}\}.
   \label{drroots}
\end{equation}
\item
If a representation \({\cal R}\) contains  a vector \(\mu\) and
its negative \(-\mu\) at the same time, then we have Ts\((L^{odd})=0\).
In such a case the corresponding Lie algebra
representations are called real.
In order to construct the odd power conserved quantities appearing in
\(A_r\) for all \(r\), \(D_r\) for odd
\(r\),
\(E_6\) and \(I(m)\) for odd \(m\), we need  a Lax pair in non-real
representations.
For \(A_r\) all the fundamental representations corresponding to the
fundamental weights \(\lambda_j\), \(j=1,\ldots,r\)
except for the middle one
\(\lambda_{(r+1)/2}\) for odd \(r\) are non-real.
 For  \(D_r\) with odd
\(r\),  the spinor and anti-spinor
representations are non-real. For \(E_6\) the {\bf 27} and \({\bf
\overline{27}}\) are non-real.
\(I_2(m)\) is the symmetry group of a regular \(m\)-sided polygon.
The set of \(m\) vectors with `half" angles
of the roots (see (\ref{dihedroots}))
to be denoted by \(V_m\) (\ref{vmpara}), provides a non-real representation
when \(m\) is odd.
\item
In Appendix B we list for each root system how the full set of independent
conserved quantities are obtained by choosing proper representations of the
Lax pair.

\end{enumerate}

\subsection{Rational potential with harmonic force}
\label{harmL}
The quantum Lax pair for the  type V models needs a separate formulation.
The
explicit form of the Hamiltonian is
\begin{equation}
   {\cal H}={1\over2}p^2+{1\over2}\omega^2q^2+
   {1\over2}\sum_{\rho\in\Delta_+}
      {g_{|\rho|}(g_{|\rho|}-1)}{|\rho|^{2}\over{(\rho\cdot q)^2}}
    -{\cal E}_0.
   \label{1stratharm}
\end{equation}
The canonical equations of motion are equivalent to the following
Lax equations for \(L^{\pm}\):
\begin{equation}
   {d\over dt}{L^\pm}=i[{\cal H},L^\pm]=
   [L^{\pm},{M}]\pm i\omega L^{\pm},
   \label{omegaLM}
\end{equation}
in which (see section 4 of \cite{bcs2})
\({M}\) is the same as before (\ref{Mtildef}), and
 \(L^{\pm}\) and \(Q\) are defined by
\begin{equation}
   L^{\pm}=L\pm i\omega Q, \quad Q=q\cdot\hat{H},
\label{Lpmdef}
\end{equation}
with \(L\), \(\hat{H}\) as earlier (\ref{LaxOpDef}), (\ref{sHdef}).
If we define hermitian operators \({\cal L}_1\) and \({\cal L}_2\) by
\begin{equation}
   {\cal L}_1=L^+L^-,\quad {\cal L}_2=L^-L^+,
   \label{defcalL}
\end{equation}
they satisfy Lax-like equations
   \begin{equation}
   \dot{{\cal L}}_k=[{\cal L}_k,{M}],\quad k=1,2.
\end{equation}
\noindent From these we can construct conserved quantities
\begin{equation}
   \mbox{Ts}({\cal L}_j^n),\quad j=1,2,\quad n=1,2,\ldots,
\end{equation}
as before.
Such quantum conserved quantities have been previously reported for
models based on \(A_r\) root systems \cite{ShasSuth,UjWa}.
It should be remarked that {Ts}\(({\cal L}_2^n)\) is no longer the same as
{Ts}\(({\cal L}_1^n)\) due to quantum corrections.
It is elementary to check that the first conserved quantities
give the Hamiltonian (\ref{1stratharm})
\begin{equation}
  {\cal H} \propto\mbox{Ts}({\cal L}_1)= \mbox{Ts}({\cal
   L}_2) +const.
\end{equation}
This then completes the presentation of the quantum Lax pairs
and quantum conserved quantities for all of
the quantum Calogero-Moser models with non-elliptic potentials.

%%%%%%%%%%%%%%%%%%%
\section{Algebraic construction of excited states I}
\label{alconst1}
\setcounter{equation}{0}
In this section we show that all the excited states of the
type V Calogero-Moser
models
can be constructed algebraically.
Later in section \ref{alconst2} we show the same results in terms of
the \(\ell\) operators to be introduced in section \ref{lops}.
The main result is surprisingly simple and can be stated universally:

Corresponding to each partition of an integer \(N\)
which specify the energy level
(\ref{omespec}) into the sum of the degrees of Coxeter
invariant polynomials
(\ref{Ndiv}), we have an eigenstate of the Hamiltonian \(\hat{\cal H}\)
with eigenvalue $\omega N+{\cal E}_0$:
\begin{equation}
   \prod_{j=1}^r\,(B_{f_j}^+)^{n_{j}}\,e^W,\quad
   N=\sum_{j=1}^rn_j f_j,\quad n_j\in{\bf Z}_+,
   \label{omeigenst}
\end{equation}
in which the integers \(\{f_j\}\), \(j=1,\ldots,r\) are listed in Table II.
They exhaust all the excited states.
In other words the above states give the complete basis of the Fock space.
The creation operators \(B_{f_j}^+\)
 and the corresponding annihilation operators
\footnote{We adopt the notation by Olshanetsky and Perelomov
\cite{OP2,Pere1}.}
 \(B_{f_j}^-\)
are defined in terms of the Lax operators \(L^\pm\)
(\ref{Lpmdef}) as follows:
\begin{equation}
   B_{f_j}^{\pm}=\mbox{Ts}(L^{\pm})^{f_j},\quad
   j=1,\ldots,r.
   \label{Bcrandef}
\end{equation}
They are hermitian conjugate to each other
\begin{equation}
   (B_{f_j}^\pm)^\dagger=B_{f_j}^\mp
\end{equation}
with respect to the standard hermitian inner product
of the states defined in \(PW\):
\begin{equation}
   (\psi,\varphi)=\int_{PW}\psi^*(q)\varphi(q)\,dq.
   \label{scalprodef}
\end{equation}
We will show later in section \ref{lops}, (\ref{cccommu}) that the
creation (annihilation) operators commute among themselves:
\begin{equation}
[B_k^+,B_l^+]=[B_k^-,B_l^-]=0,\quad k,l\in\{f_j|\ j=1,\ldots,r\},
\end{equation}
so that the state (\ref{omeigenst}) does not
depend on the order of the creation.

\bigskip
The proof is very simple.
By using (\ref{omegaLM}) we obtain
\begin{equation}
   {d\over dt}{(L^\pm)^n}=i[{\cal H},(L^\pm)^n]
   =[(L^{\pm})^n,{M}]\pm in\omega (L^{\pm})^n,
   \label{nomegaLM}
\end{equation}
from which
\begin{equation}
   [{\cal H}, B_n^\pm]=\pm n\omega B_n^\pm,
\end{equation}
follows after taking the total sum.
This simply says that \(B_n^\pm\) creates (annihilates)
a state having energy \(n\omega\).
In other words we have
\[
   \hat{\cal H}\prod_{j=1}^r\,(B_{f_j}^+)^{n_{j}}\,e^W
   =\left({\cal E}_0+\omega
   \sum_{k=1}^rn_k f_k\right)\prod_{j=1}^r\,(B_{f_j}^+)^{n_{j}}\,e^W.
\]
Moreover, it is trivial to show that
\begin{equation}
   \sum_{\nu\in{\cal R}}(L^-)_{\mu\nu}\,e^W=
    \left(p\cdot\mu-i\omega q\cdot\mu
   +i\sum_{\rho\in\Delta_+}{\rho\cdot\mu\over{\rho\cdot q}}\right)\,e^W=
   \mu\cdot\!\left(p+i{\partial W\over{\partial q}}\right)e^W=0,
\end{equation}
which implies that the ground state is annihilated by all the
annihilation operators
\begin{equation}
   B_{f_j}^-\,e^W=0,\quad j=1,\ldots,r.
\end{equation}

Some remarks are in order.
\begin{enumerate}
\item
In most cases the energy levels are highly degenerate.
The above basis is neither
orthogonal nor normalised.
\item
The independence of the creation-annihilation operators can also be shown
in a similar way to that of the conserved quantities.
 As with the
conserved quantities,  plural representations are  necessary to define
the full set of creation-annihilation operators in some models.
This aspect will be discussed in later sections in
connection with the \(\ell\) operators.
\item
Reflecting the universality of the first exponent, \(f_1=2\),
the creation and annihilation operators of the
least quanta, \(2\omega\), exist in all the models.
They form an \(sl(2,{\bf R})\) algebra
together with the Hamiltonian \(\hat{\cal H}\):
\begin{equation}
   [\hat{\cal H},b^\pm_2]=\pm2\omega b^\pm_2,\quad
   [b_2^+,b_2^-]=-\omega^{-1}\hat{\cal H},
\end{equation}
in which \(b_2^\pm\) are  normalised forms of \(B_2^\pm\):
\begin{equation}
   b_2^\pm=\sum_{\mu,\nu\in{\cal R}}(L^\pm)^2_{\mu\nu}/(4\omega C_{\cal
   R}).
\label{normb2def}
\end{equation}
 The \(sl(2,{\bf R})\) algebra was discussed by many authors
(see, for example, \cite{Pere1,Gamb,Br,Heck2} and others) in connection
with  the models based on classical root systems. We will show later in
subsection
\ref{B2plus} that the
states created by
\(B_2^+\) (\(b_2^+\)) only can be expressed by  the Laguerre polynomial:
\begin{equation}
   (b_2^+)^n\,e^W=n! L^{(\tilde {\cal E}_0-1)}_n(\omega q^2)~e^W,\quad
   \tilde {\cal E}_0\equiv{\cal E}_0/\omega.
   \label{B2series}
\end{equation}
It is trivial to verify that \(L^{(\tilde {\cal E}_0-1)}_n(\omega q^2)\)
is an eigenfunction of \(\tilde{\cal H}\) (\ref{Htilom})
\begin{equation}
   \tilde{\cal H}L^{(\tilde {\cal E}_0-1)}_n(\omega q^2)=
   2n\omega L^{(\tilde
   {\cal E}_0-1)}_n(\omega q^2).
\end{equation}
The normalisation of the state
\begin{equation}
   |\!|(b_2^+)^n\,e^W|\!|^2=n!{\cal N}_0/\Gamma(n+\tilde
   {\cal E}_0),\quad {\cal N}_0\equiv|\!|e^W|\!|^2\Gamma(\tilde
   {\cal E}_0),
\end{equation}
is also dictated by the \(sl(2,{\bf R})\) relations.
The Laguerre polynomial wavefunctions appear as `radial' wavefunctions
in all the cases \cite{Cal1}.
This will be shown explicitly for
for the rank two models given in subsection \ref{ranktwo}.
\item
As is emphasised by Perelomov \cite{Pere1} and Gambardella \cite{Gamb}
the \(sl(2,{\bf R})\) algebra and the corresponding
Laguerre wavefunctions
are more universal than Calogero-Moser models.
They arise when the potentials are homogeneous functions
in \(q\) of degree
\(-2\) with the confining harmonic force.
\item
The operators \(\{Q_n\}\) and \(\{B_n^\pm\}\) do not form a Lie algebra.
They satisfy interesting non-linear relations, for example,
\begin{equation}
   [[B_n^+,b_2^-],b_2^+]=nB_n^+,\quad
   [[B_n^-,b_2^+],b_2^-]=nB_n^-.
   \label{highalge}
\end{equation}
This tells, for example, that although \(B_n^+\) and \(b_2^+\)
create different units of quanta \(n\) and \(2\), they are not independent
\[
   [B_n^+,b_2^-]\neq0\neq [B_n^-,b_2^+].
\]
Clarification of the algebraic structure (\ref{highalge}) for each root
system is wanted.
\end{enumerate}

%%%%%%%%%%%%%%%%%%%
\section{\(\ell\) operators}
\label{lops}
\setcounter{equation}{0}
In this section we will show the equivalence of the
quantum conserved quantities obtained in the Lax operator formalism
of section \ref{cons} and those derived in the `commuting differential
operators' formalism initiated by Dunkl \cite{Dunk} and followed by many
authors.
Again the equivalence is universal, applicable to the models based on
any root systems.
We propose to call the operators in the latter approach simply
`\(\ell\) operators', since they are essentially the same as the
\(L\) operator in the Lax pair formalism and that they are not mutually
commuting, as we will show presently, when the interaction potentials are
trigonometric (hyperbolic), (\ref{llcomm}).
Although these two formalisms are formally equivalent, the \(\ell\)
operator formalism has many advantages over the Lax pair
one. Roughly speaking, the `vector-like' objects \(\ell_\mu\)'s are easier
to handle than the matrix \(L_{\mu\nu}\).

Let us fix a representation \({\cal R}\) of the Coxeter group
\(G_{\Delta}\)
and define for each element \(\mu\in{\cal R}\) the following
differential-reflection operator
\begin{equation}
   \ell_{\mu}=\ell\cdot\mu=p\cdot\mu+i
   \sum_{\rho\in\Delta_{+}}g_{|\rho|}
   \,(\rho\cdot\mu)\,x(\rho\cdot q)\check{s}_{\rho},\quad \mu\in{\cal R}.
   \label{ldefs}
\end{equation}
It is linear in \(\mu\) and Coxeter covariant
\begin{equation}
   \ell_{\mu+\nu}=\ell_{\mu}+\ell_{\nu},\qquad
   \check{s}_{\rho}\ell_{\mu}\check{s}_{\rho}=\ell_{s_{\rho}(\mu)},\quad
   \forall \rho\in\Delta.
\end{equation}
They are hermitian operators, \(\ell_{\mu}^\dagger=\ell_{\mu}\),
with respect to
the standard inner product
for  the states (\ref{scalprodef}).

It is straightforward to show that the quantum conserved quantities
\(Q_n\) derived in the previous section (\ref{qconv}) can be expressed
as polynomials in
 the \(\ell\) operators as follows:
\begin{equation}
   Q_n\psi=\sum_{\mu,\nu\in{\cal R}}(L^n)_{\mu\nu}\psi=
   (\sum_{\mu\in{\cal R}}\ell_{\mu}^n)\psi,
   \label{Llequal}
\end{equation}
in which \(\psi\) is an arbitrary Coxeter invariant state,
\(\check{s}_{\rho}\psi=\psi\).
This also illustrates the Coxeter invariance of \(Q_n\) clearly, since
\(\check{s}_{\rho}(\sum_{\mu\in{\cal R}}\ell_{\mu}^n)\check{s}_{\rho}
=\sum_{\mu\in{\cal R}}\ell_{{s}_{\rho}(\mu)}^n=\sum_{\mu\in{\cal
R}}\ell_{\mu}^n\). For \(n=1\) it is trivial, since
\begin{eqnarray}
   \sum_{\nu\in{\cal R}}(L)_{\mu\nu}\psi&=&
   \left(p\cdot\mu+i
   \sum_{\rho\in\Delta_{+}}g_{|\rho|}
      \,(\rho\cdot\mu)\,x(\rho\cdot
   q)\sum_{\nu\in{\cal R}}(\hat{s}_{\rho})_{\mu\nu}\right)\psi
   \nonumber\\
   &=&\left(p\cdot\mu+i
   \sum_{\rho\in\Delta_{+}}g_{|\rho|}
      \,(\rho\cdot\mu)\,x(\rho\cdot
   q)\check{s}_{\rho}\right)\psi=\ell_{\mu}\psi,
\end{eqnarray}
in which \(\sum_{\nu\in{\cal R}}(\hat{s}_{\rho})_{\mu\nu}=1\) and
\(\check{s}_{\rho}\psi=\psi\) are used.
Let us assume that
\begin{equation}
\sum_{\nu\in{\cal R}}(L^n)_{\mu\nu}\psi=
\ell_{\mu}^n\psi,
\end{equation}
is correct, then we obtain
\begin{eqnarray*}
   &&\sum_{\nu\in{\cal R}}(L^{n+1})_{\mu\nu}\psi=
   \sum_{\lambda, \nu\in{\cal R}}L_{\mu\lambda}(L^{n})_{\lambda\nu}\psi=
   \sum_{\lambda\in{\cal
   R}}L_{\mu\lambda}\ell_{\lambda}^n\psi,\\
   &=&\sum_{\lambda\in{\cal R}}\left(
   p\cdot\mu\delta_{\mu\lambda}+i
   \sum_{\rho\in\Delta_{+}}g_{|\rho|}
      \,(\rho\cdot\mu)\,x(\rho\cdot
   q)(\hat{s}_{\rho})_{\mu\lambda}
   \right)\ell_{\lambda}^n\psi.
\end{eqnarray*}
In the second summation only such \(\lambda\) as \(\lambda=s_{\rho}(\mu)\)
contributes and we find
\[
   \ell_{s_{\rho}(\mu)}^n\psi=(\check{s}_{\rho}
   \ell_{\mu}^n\check{s}_{\rho})\psi
   =\check{s}_{\rho}\ell_{\mu}^n\psi.
\]
Thus we arrive at
\begin{equation}
   \sum_{\nu\in{\cal R}}(L^{n+1})_{\mu\nu}\psi=
   \ell_{\mu}^{n+1}\psi,
\end{equation}
and the equivalence of the two expressions of the conserved
quantity (\ref{Llequal}) is proved.

Commutation relations among \(\ell\) operators can be evaluated in
a similar manner as those appearing in the Lax pair \cite{bcs2,bms},
that is, by
decomposing the roots into two-dimensional sub-root systems.
We obtain
\begin{equation}
   [\ell_{\mu},\ell_{\nu}]=-a^2\sum_{\rho,\sigma\in\Delta_{+}}
   g_{|\rho|}g_{|\sigma|}
   \,(\rho\cdot\mu)\,(\sigma\cdot\nu)[\check{s}_{\rho},\check{s}_{\sigma}]
   \times\left\{
   \begin{array}{rl}
      0&\mbox{rational},\\
      -1&\mbox{hyperbolic},\\
      1&\mbox{trigonometric}.
   \end{array}
   \right.
   \label{lcomrel}
\end{equation}
One important use of the \(\ell\) operators is the proof of
involution of quantum
conserved quantities.
For type I models Heckman \cite{Heck2} gave a universal proof
based on the commutation relation (\ref{lcomrel}):
\begin{equation}
   [Q_n,Q_m]\psi=
   \sum_{\mu,\nu\in{\cal R}}[\ell_\mu^n,\ell_\nu^m]\psi=0,\quad
   \mbox{rational model}.
\end{equation}
This was the motivation for the introduction of the commuting
differential-reflection operators by Dunkl, \cite{Dunk}.
In fact, Dunkl's and Heckman's operators were the similarity
transformation of
\(\ell_\mu\) by the ground state wavefunction \(e^W\):
\begin{equation}
   \tilde{\ell}_\mu=e^{-W}\ell_\mu e^W=
   p\cdot\mu+i
   \sum_{\rho\in\Delta_{+}}g_{|\rho|}
      \,{(\rho\cdot\mu)\over (\rho\cdot q)}(\check{s}_{\rho}-1).
\end{equation}

As for type V  models, we define
\(\ell^{\pm}\) corresponding to \(L^{\pm}\) (\ref{Lpmdef}):
\begin{equation}
   \ell_{\mu}^{\pm}=\ell^\pm\cdot\mu=p\cdot\mu \pm i\omega(q\cdot\mu)+i
   \sum_{\rho\in\Delta_{+}}g_{|\rho|}
       \,{(\rho\cdot\mu)\over (\rho\cdot q)}\check{s}_{\rho},\quad
   \mu\in{\cal R}.
   \label{lpmdefs}
\end{equation}
They are linear in \(\mu\), Coxeter covariant and hermitian conjugate of
each other with respect to the standard inner product (\ref{scalprodef}):
\begin{equation}
   \check{s}_{\rho}\ell_{\mu}^\pm\check{s}_{\rho}=
   \ell_{s_{\rho}(\mu)}^\pm,\qquad (\ell_{\mu}^\pm)^\dagger=
   \ell_{\mu}^\mp.
\end{equation}

The conserved quantities are expressed as polynomials in \(\ell^{\pm}\)
operators:
\begin{eqnarray}
   \mbox{Ts}({\cal L}_1^n)\psi&=&\sum_{\mu,\nu\in{\cal
   R}}(L^+L^-)^n_{\mu\nu}\psi=
   \sum_{\mu\in{\cal R}}(\ell_{\mu}^+\ell_{\mu}^-)^n\psi,
   \label{lpmcons}\\
   \mbox{Ts}({\cal L}_2^n)\psi&=&\sum_{\mu,\nu\in{\cal
   R}}(L^-L^+)^n_{\mu\nu}\psi=
   \sum_{\mu\in{\cal R}}(\ell_{\mu}^-\ell_{\mu}^+)^n\psi.
   \nonumber
\end{eqnarray}
Likewise the creation and annihilation operators \(B_n^\pm\)
 (\ref{Bcrandef})
are expressed as
\begin{equation}
   B_n^\pm\psi=\mbox{Ts}(L^{\pm})^n\psi=\sum_{\mu,\nu\in{\cal
   R}}(L^\pm)_{\mu\nu}\psi=
   \sum_{\mu\in{\cal R}}(\ell_{\mu}^\pm)^n\psi.
\end{equation}

The commutation relations among \(\ell^\pm\) operators
are easy to evaluate, since
 \(\ell\) operators commute in the rational potential models
(\ref{lcomrel}):
\begin{equation}
   [\ell_{\mu}^+,\ell_{\nu}^+]=[\ell_{\mu}^-,\ell_{\nu}^-]=0,\quad
   [\ell^-_{\mu},\ell^+_{\nu}]=2\omega\left(\mu\cdot\nu+
   \sum_{\rho\in\Delta_{+}}g_{|\rho|}
   \,(\rho\cdot\mu)\,(\rho^\vee\!\cdot\nu)\check{s}_{\rho}\right).
   \label{llcomm}
\end{equation}
\noindent From these it follows that the creation (annihilation) operators
\(B_n^\pm\) do commute among themselves:
\begin{equation}
   [B_n^+,B_m^+]\psi=[B_n^-,B_m^-]\psi=0.
   \label{cccommu}
\end{equation}
It is also clear that \(\ell_\mu^\pm{/\sqrt{2\omega}}\)
are the `deformation' of the creation (annihilation) operators of  the
ordinary multicomponent harmonic oscillators. In fact we have
\begin{equation}
   \ell_\mu^+~e^{W}=2i\omega(\mu\cdot q)~e^W \quad{\rm
   and}\quad \ell_\mu^-~e^{W}=0.
\end{equation}
In the next section we present an alternative scheme of algebraic
construction of excited states of type V models by pursuing the analogy that
\(\ell^\pm\) are the creation and annihilation operators of the unit
quantum.
This method was applied to the \(A_r\) models by Brink et. al and others
\cite{Br,Pol,UjWa}.

%%%%%%%%%%%%%%%%%%%
\section{Algebraic construction of excited states II}
\label{alconst2}
\setcounter{equation}{0}

\subsection{Operator solution of the triangular Hamiltonian}
\label{PNcreationop}
In subsection \ref{harmtri}, we have shown that an eigenfunction of
\({\cal H}\) with eigenvalue \(N\omega\) is given by
\begin{equation}
\left(P_N(q)+\tilde{P}_{N-2}(q)\right)e^W,
\end{equation}
in which \(P_N(q)\) is a Coxeter invariant polynomial in \(q\)
of homogeneous
degree \(N\) and \(\tilde{P}_{N-2}(q)\) is a Coxeter invariant
polynomial in
\(q\) of degree \(N-2\) and lower. The non-leading polynomial
\(\tilde{P}_{N-2}(q)\) is completely determined by the leading one
\(P_N(q)\) due to the triangularity.
This solution can be written in an operator form as follows.

Suppose \(P_N(q)\) is expressed as
\begin{equation}
   P_N(q)=\sum_{\{\mu\}}c_{\{\mu\}}(q\cdot\mu_1)\cdots(q\cdot\mu_N),
   \quad \mu_j\in{\cal R},\quad c_{\{\mu\}}:\ const.
\end{equation}
We obtain a Coxeter invariant polynomial
in the creation operators \(\ell^+\)
by replacing \(q\cdot\mu\) by \(\ell^+_\mu/(2i\omega)\):
\[
   P_N(q)\Rightarrow {1\over{(2i\omega)^N}}P_N(\ell^+).
\]
This creates the above eigenfunction of \({\cal H}\) from the ground
state:
\begin{equation}
   {1\over{(2i\omega)^N}}P_N(\ell^+)\,e^W=
   \left(P_N(q)+\tilde{P}_{N-2}(q)\right)e^W.
   \label{pncreate}
\end{equation}

The proof is again elementary.
By using the commutation relations among
\(\ell^\pm\) operators it is straightforward to derive the explicit
expression of the Hamiltonian  in terms of \(\ell^\pm\):
\begin{equation}
   {\cal
   H}=\frac{1}{2\mathcal{C_R}}
   \sum_{\mu\in\mathcal{R}}\ell^+_{\mu}\ell^-_{\mu}
   +
   \sum_{\rho\in\Delta_{+}}g_{|\rho|}\left(\omega
   +\frac{1}{2}{|\rho|^2\over{(\rho\cdot q)^2}}\right)(\check{s}_{\rho}-1),
   \label{llham}
\end{equation}
in which the second term vanishes upon acting on a
Coxeter invariant state.
Next we obtain
\begin{equation}
   \frac{1}{2\mathcal{C_R}}
   \sum_{\mu\in\mathcal{R}}[\ell^+_{\mu}\ell^-_{\mu},
   \ell_\nu^\pm]=[\ell_\nu^\pm,S]\pm\omega\ell_\nu^\pm,\quad
   S\equiv\sum_{\rho\in\Delta_+}g_{|\rho|}\check{s}_{\rho},
\end{equation}
which is an \(\ell\) operator version of (\ref{omegaLM}).
Since a commutator is a derivation, we obtain
\begin{equation}
   \frac{1}{2\mathcal{C_R}}
   \sum_{\mu\in\mathcal{R}}[\ell^+_{\mu}\ell^-_{\mu},
   P_N(\ell^+)]=[P_N(\ell^+),S]+N\omega P_N(\ell^+),
\end{equation}
in which the first term in r.h.s. vanishes due to the Coxeter invariance
of \(P_N\). Thus we arrive at the desired commutation relation
\begin{equation}
   [{\cal H}, P_N(\ell^+)]=N\omega P_N(\ell^+)+
   \sum_{\rho\in\Delta_{+}}g_{|\rho|}\left[
   \frac{1}{2}{|\rho|^2\over{(\rho\cdot
   q)^2}},P_N(\ell^+)\right](\check{s}_{\rho}-1),
\end{equation}
and the eigenvalue equation
\begin{equation}
   {\cal H}\,P_N(\ell^+)\,e^W=N\omega P_N(\ell^+)\,e^W.
\end{equation}
Since the action of the creation operators on the ground state is
\begin{equation}
   \ell_{\mu_1}^+\cdots\ell_{\mu_N}^+\,e^W=[
   (2i\omega)^N(q\cdot\mu_1)\cdots(q\cdot\mu_N)
   +\mbox{lower powers of } q]\,e^W,
\end{equation}
our assertion (\ref{pncreate}) is proved.
It should be stressed that in this formalism the Coxeter invariance of the
polynomial \(P\) is important but not how it is obtained.

Like the above Hamiltonian (\ref{llham}),  the \(\ell\) operator formulas
of higher conserved quantities (\ref{lpmcons}) contain extra terms:
\begin{equation}
   \mbox{Ts}({\cal L}_1^n)=\sum_{\mu,\nu\in{\cal
   R}}(L^+L^-)^n_{\mu\nu}=
   \sum_{\mu\in{\cal R}}(\ell_{\mu}^+\ell_{\mu}^-)^n+ VT.
\end{equation}
Here \(VT\) stands for vanishing terms when they act on a Coxeter invariant
state. The same is true for most formulas derived in section
\ref{lops}.

\subsection{States Created by \(B_2^+\)}
\label{B2plus}

Here we derive the explicit forms of the subseries of eigenstates obtained
by multiple applications of the least quanta creation operator \(B_2^+\)
(\ref{Bcrandef}), or its normalised form \(b_2^+\) (\ref{normb2def}).
It is convenient to work with the similarity transformed operator
\begin{equation}
   \tilde{b}_2^+=e^{-W}b_2^+\,e^W={1\over{4\omega C_{\cal
   R}}}\sum_{\mu\in{\cal R}}(\tilde{\ell}_{\mu}^+)^2+VT,
\end{equation}
in which
\begin{equation}
   \tilde{\ell}_{\mu}^+=
   p\cdot\mu+2i\omega(q\cdot\mu)+i\sum_{\rho\in\Delta_+}
   {\rho\cdot\mu\over{\rho\cdot q}}(\check{s}_{\rho}-1).
\end{equation}
Let \(f(u)\) be an arbitrary function of \(u\equiv\omega q^2\), then it is
Coxeter invariant.
We find
\begin{equation}
   \tilde{\ell}_{\mu}^+f(u)=2i\omega(q\cdot\mu)(1-{d\over{du}})f(u),\quad
   u\equiv\omega q^2,
\end{equation}
and
\begin{equation}
   \tilde{b}_2^+f(u)=-\left[u\left(1-{d\over{du}}\right)^2
   -\tilde{\cal E}_0\left(1-{d\over{du}}\right)\right]f(u).
\end{equation}
Since \(\tilde{b}_2^+1=\tilde{\cal E}_0-u=L_1^{(\tilde{\cal E}_0-1)}(u)\),
we
assume
\begin{equation}
   (\tilde{b}_2^+)^n1=n!L_n^{(\tilde{\cal E}_0-1)}(u).
\end{equation}
By using the Laguerre differential equation (\ref{lagueeq})
and the recurrence formulas of
the Laguerre polynomial \(L_n^{(\alpha)}(u)\),
\begin{eqnarray}
   &&u{d\over{du}}L^{(\alpha)}_n(u)=nL^{(\alpha)}_n(u)
   -(n+\alpha)L^{(\alpha)}_{n-1}(u),\\
   &&nL^{(\alpha)}_n(u)+(u-2n-\alpha+1)L^{(\alpha)}_{n-1}(u)
   +(n+\alpha-1)L^{(\alpha)}_{n-2}(u)=0,
\end{eqnarray}
we can show
\begin{equation}
   -\left[u(1-{d\over{du}})^2-\tilde{\cal E}_0(1-{d\over{du}})
   \right]L_n^{(\tilde{\cal
   E}_0-1)}(u)=(n+1)L_{n+1}^{(\tilde{\cal E}_0-1)}(u).
\end{equation}
Thus the induction is proved and we arrive at (\ref{B2series}).
The orthogonality of the states
\begin{equation}
   \left((B_2^+)^ne^W,(B_2^+)^me^W\right)=0,\quad n\neq m
\end{equation}
can be easily understood as the \(du\) part of the measure
\[
   e^{2W}d^r\!q= e^{-u}u^{\tilde{\cal E}_0-1}dud\Omega,
   \quad d\Omega: \mbox{angular
   part},
\]
is the proper weight function for the Laguerre polynomial
\(L_n^{(\tilde{\cal E}_0-1)}(u)\).

\subsection{Explicit solutions of the rank two models}
\label{ranktwo}
For rank two models, the Liouville integrability, or the involution of
conserved quantities is automatically satisfied since the second conserved
quantity is already obtained.
For rank two type V models, the complete set of orthogonal wavefunctions
can be  written down explicitly in terms of separation of variables by
using the Coxeter invariant polynomials.
These are based on the dihedral root systems \(I_2(m)\), with
\(A_2\cong I_2(3)\) \cite{Cal1},
\(B_2\cong I_2(4)\) and \(G_2\cong I_2(6)\) \cite{G2sep}.
The Coxeter invariant polynomials exist at degree 2, {\em i.e.\/}
\(q^2\) and \(m\)
which is
\begin{equation}
   \prod_{j=1}^m(v_j\cdot q),
\end{equation}
where \(\{v_j\}\) is a set of vectors given in (\ref{vmpara}).
If we introduce the two-dimensional polar coordinates system
\footnote{We believe no confusion arises here, between the radial coordinate
variable \(r\) and the rank of the root system \(r\), which in this case is
2
of \(I_2(m)\).}
for \(q\)
\begin{equation}
   q=r(\sin\theta,\cos\theta),
\end{equation}
then the principal Weyl chamber is
\begin{equation}
   PW:\ 0<r^2<\infty,\quad 0<\theta<\pi/m.
\end{equation}
The two Coxeter invariant variables read:
\begin{equation}
   q^2=r^2,\quad \prod_{j=1}^m(v_j\cdot q)=2({r\over2})^m\cos m\theta,
\end{equation}
and the latter variable varies the full range, \(-1<\cos m\theta<1\)
in the \(PW\).
Thus solving the eigenvalue equation for \(\tilde{\cal H}\)
(\ref{htileigen}) by
separation of variables in the polar coordinate system is
compatible with Coxeter
invariance.
We adopt as two independent variables
\begin{equation}
   u\equiv\omega r^2,\quad z\equiv\cos m\theta.
   \label{coxinvvartwo}
\end{equation}
The solutions consist of a Gegenbauer (Jacobi) polynomial in
\(\cos m\theta\)
times a Laguerre polynomial in \(\omega r^2\).
The former we have encountered in the \(A_1\) Sutherland problem, subsection
\ref{triginvcase} and the latter in the \(A_1\) Calogero problem subsections
\ref{harmtri} and \ref{B2plus}.

Let us demonstrate this for odd \(m\) with a single coupling constant and
for
even \(m\) with two independent coupling constants, in parallel.
In terms of the Coxeter invariant variables (\ref{coxinvvartwo}) the
\(I_2(m)\) Hamiltonians take surprisingly simple forms:
\begin{eqnarray}
   \tilde{\cal H}&=&\omega r{\partial\over{\partial
   r}}-{1\over2r}{\partial\over{\partial r}}
   \left(r{\partial\over{\partial r}}\right)
   -{1\over2r^2}\left[{\partial^2\over{\partial \theta^2}}+m{2g\cot m\theta
   \brace
   -g_0\tan{m\theta\over2}+g_e\cot{m\theta\over2}}{\partial\over{\partial
   \theta}}\right]\\[10pt]
   &=&-2\omega\left[u{\partial^2\over{\partial u^2}}
  +(1-u){\partial\over{\partial
   u}}\right] -{\omega m^2\over{2u}}
   \left[(1-z^2){\partial^2\over{\partial z^2}}+
   {0\brace g_o-g_e} -{1+2g\brace 1+g_e+g_o}z{\partial\over{\partial
   z}}\right].\nonumber
\end{eqnarray}
The \(z\) part admits polynomial solutions
\begin{eqnarray}
   \left[(1-z^2){d^2\over{dz^2}}+
   {0\brace g_o-g_e}
   -{1+2g\brace
   1+g_e+g_o}z{d\over{dz}}\right]&&\hspace{-.5cm}
   P_{\ell}^{{(g-{1\over2},g-{1\over2})
  \brace(g_o-{1\over2},
   g_e-{1\over2})}}(z)
   \nonumber\\[10pt]
   =-\ell\left(\ell+{2g\brace g_e+g_o}\right)
   &&\hspace{-.5cm}
   P_{\ell}^{{(g-{1\over2},g-{1\over2})
   \brace(g_o-{1\over2},g_e-{1\over2})}}(z),
\end{eqnarray}
in which \(\ell\) is the degree of the polynomial.
After substituting them, the radial part of the Hamiltonian
\(\tilde{\cal H}_r\)
reads
\begin{equation}
   \tilde{\cal H}_r=-2\omega\left[u{d^2\over{du^2}}
   +(1-u){d\over{du}}
-{m^2\over{4u}}\ell\left(\ell+{2g\brace g_e+g_o}\right)\right].
\end{equation}
By similarity transformation in terms of \(u^{m\ell/2}\propto r^{m\ell}\),
which is
the radial part of the highest term of the
polynomial \(P^{(\alpha,\beta)}_{\ell}(r^m\cos m\theta)\), it reads
\begin{equation}
   u^{-m\ell/2}\tilde{\cal H}_ru^{m\ell/2}=
   -2\omega\left[u{d^2\over{du^2}}+\left(m\left({\ell +g\brace
   \ell+{1\over2}(g_e+g_o)}\right)+1-u\right)
   {d\over{du}}-{m\ell\over2}\right].
\end{equation}
This is the main part of the differential equation for
the Laguerre polynomial
(\ref{lagueeq}):
\[
   \left[u{d^2\over{du^2}}+\left(m\ell+\tilde{\cal
   E}_0-u\right){d\over{du}}-n\right]L^{(m\ell+\tilde{\cal
   E}_0-1)}_n(u)=0,
\]
in which the indices \(m(\ell+g)\) and \(m(\ell+(g_o+g_e)/2)\) can be
written
in a unified way as
\(m\ell+\tilde{\cal E}_0-1\).
Thus the eigenstates of the Hamiltonian are obtained:
\begin{eqnarray}
   \tilde{\cal H}_ru^{m\ell/2}L^{(m\ell+\tilde{\cal
   E}_0-1)}_n(u)\hspace{-.3cm}&=&\hspace{-.3cm}
   \omega(2n+m\ell)u^{m\ell/2}L^{(m\ell+\tilde{\cal
   E}_0-1)}_n(u),\\[12pt]
   \hat{\cal H}u^{m\ell/2}L^{(m\ell+\tilde{\cal
   E}_0-1)}_n(u)\hspace{-.3cm}&&\hspace{-.7cm}
   P_{\ell}^{{(g-{1\over2},g-{1\over2})
   \brace(g_o-{1\over2},g_e-{1\over2})}}\!(z)\\
   &=&\left(\omega(2n+m\ell)\!+\!{\cal E}_o\right)
   \!u^{m\ell/2}L^{(m\ell+\tilde{\cal
   E}_0-1)}_n(u)P_{\ell}^{{(g-{1\over2},g-{1\over2})
   \brace(g_o-{1\over2},g_e-{1\over2})}}\!(z).
   \nonumber
\end{eqnarray}

It is instructive to note that the Hamiltonians \(\hat{\cal H}\)
look also simple:
\begin{equation}
   \hat{\cal H}=\omega r{\partial\over{\partial
   r}}
   -{1\over2r}{\partial\over{\partial r}}
   \left(r{\partial\over{\partial r}}
   \right)
   -{1\over2r^2}{\partial^2\over{\partial\theta^2}}
   -{m^2\over2r^2}\left\{\begin{array}{c}
   {g(g-1)\over{\sin^2m\theta}}\\[12pt]
   {g_o(g_o-1)\over{4\cos^2{m\theta\over2}}}+
   {g_e(g_e-1)\over{4\sin^2{m\theta\over2}}}
   \end{array}
   \right\}.
\end{equation}
Olshanetsky and Perelomov \cite{OP2} obtained the above solutions
starting from these formulas.

\section{Involution of conserved quantities}
\label{invopr}
\setcounter{equation}{0}

\subsection{Universal proof of involution of quantum conserved quantities
for
type I, II and III models}
\label{uniproof}
Here we present a proof of involution of quantum conserved quantities
\(\{Q_n\}\)
derived from the universal Lax pair in subsection \ref{genconv}
for type I, II and III models.
The proof is applicable to all models based on any root systems.
Though a universal proof of involution for type I models is given by Heckman
\cite{Heck2} as recapitulated in section \ref{lops}, we believe the
universal
proof applicable to type II and III models as well is new.
It depends on a theorem by Olshanetsky and Perelomov \cite{OP3}.
Our own contribution is that we have provided a universal
Lax pair and conserved
quantities satisfying all the requirements of the theorem.

Liouville's theorem states the complete integrability as the existence of an
involutive set of conserved quantities as many as the degrees of freedom.
We have already given conserved quantities \(\{Q_n\}\) (\ref{qconv})
independent and as many as the degrees of freedom (see Appendix B).
They have the following properties:
\begin{enumerate}
\item
Coxeter invariance
\begin{equation}
    Q_n(s_{\rho}(p),s_{\rho}(q))=Q_n(p,q),\quad \forall\rho\in\Delta.
\end{equation}
\item
\(Q_n(p,q)\) is a homogeneous polynomial of degree \(n\) in variables
\((p_1,\ldots,p_r,x(\rho\cdot q))\).
\item
Scaling property for those of type I models:
\begin{equation}
   {}^IQ_n(\kappa^{-1}p,\kappa q)=\kappa^{-n}\,{}^IQ_n(p,q)
\end{equation}
as a consequence of the above point.
\item
For type II and III models, the asymptotic behaviour near the origin:
\begin{equation}
   Q_n(p,q)={}^IQ_n(p,q)(1+{\cal O}(|q|)),\quad \mbox{for}\quad |q|\to0.
\end{equation}
\end{enumerate}
We need to show the vanishing of
\begin{equation}
J_{lm}\equiv [Q_l,Q_m],
\end{equation}
which is a polynomial in \(\{p\}\) of degree \(s\)
\begin{equation}
   s<l+m.
   \label{slesslm}
\end{equation}
Let us decompose \(J_{lm}\) into the leading part and the rest:
\begin{equation}
   J_{lm}=J_{lm}^0+J_{lm}^{rest},\quad
   J_{lm}^0=\sum c^{j_1,\ldots,j_s}(q)p_{j_1}\ldots p_{j_s}
\end{equation}
and \(J_{lm}^{rest}\) is a polynomial in \(\{p\}\)
of degree less than \(s\).
\noindent From Jacobi identity and conservation \([{\cal H}, Q_{l(m)}]=0\),
we obtain
\begin{equation}
   [{\cal H},J_{lm}]=0.
\end{equation}
Considering the explicit form of the Hamiltonian (\ref{qCMHamiltonian})
(\(\omega=0)\), the leading ({\em i.e.\/} of degree \(s+1\) in \(\{p\}\))
part
of \([{\cal H},J_{lm}]\) comes only from the free part
\[
   [p^2,J_{lm}^0]
\]
and it vanishes if the following conditions are satisfied:
\begin{equation}
   \sum_{\sigma}{\partial\over{\partial q_t}}c^{k_1,\ldots,k_s}(q)
   =0,
\label{cqeq}
\end{equation}
where the sum is taken over all permutations of indices
\(\sigma(t,k_1,\ldots,k_s)=(j_1,\ldots,j_{s+1})\).
In \cite{Bere} it is proved (Lemma 2.5, p. 407) that the system
(\ref{cqeq}) has
only polynomial solutions.
Then Olshanetsky and Perelomov argue that for type I models the scaling
property tells that \(c^{k_1,\ldots,k_s}(\kappa
q)=\kappa^{s-l-m}c^{k_1,\ldots,k_s}(q)\). Since \(s<l+m\)
(\ref{slesslm}), it follows that the only polynomial solution  satisfying
the condition is the null polynomial. Thus we obtain
\(c^{j_1,\ldots,j_s}(q)=0\) \(\Rightarrow J_{lm}^0=0\) and \(J_{lm}=0\).
The same results follow for type II and III models by considering the
asymptotic behaviour for \(|q|\to0\). Thus the involution of all the
conserved quantities \(\{Q_n\}\) is proved. This result also implies the
involution of classical conserved quantities by taking the classical
limit (\(\hbar\to0\)).

\subsection{Rational  models with the harmonic confining force}
In this subsection we show the involution of quantum
conserved quantities  for the type V models  based on the root systems of
classical Lie algebras. The method is a straightforward generalisation of
the one developed by Polychronakos on the \(A_r\) model.
This is made possible by the availability of the universal Lax pair
formalism \cite{bcs2,bms}, in particular the root type and minimal
type Lax pairs.
We apply it to the models based on \(B_r\) and \(D_r\) root systems.
For the rational potential, the \(B_r\), \(C_r\) and \(BC_r\) models are
equivalent.
Let us choose \({\cal R}=\{\pm e_j\in{\bf R}^r|\ e_j\cdot
e_k=\delta_{jk}\}\) as the representation space of the Coxeter group
consisting of  orthogonal vectors and their negatives.
They are the set of short roots of \(B_r\) and the set of vector weights of
\(D_r\) in the parametrisation of roots given in (\ref{brroots}) and
(\ref{drroots}), respectively.

The conserved quantities in the \(\ell\) operator form are given by
\begin{equation}
   Q_n={}^{\ell}Q_n+VT,\quad
   {}^{\ell}Q_n\equiv\sum_{j=1}^r(\ell_j^+\ell_j^-)^n,\quad
   n=1,\ldots,r,
\end{equation}
in which we abbreviate \(\ell_{e_j}^\pm\) as \(\ell_{j}^\pm\).
In this case the commutation relation among \(\ell\) operators
(\ref{llcomm}) are greatly simplified thanks to the orthogonality
of
\(\{e_j\}\)'s and the explicit forms of the roots:
\begin{equation}
   [\ell_j^-,\ell_k^+]=
   -2\omega\tilde{g}(\check{s}_{jk}-\bar{\check{s}}_{jk}),
\end{equation}
in which
\begin{equation}
   \tilde{g}=\left\{
   \begin{array}{cl}
      g_L&B_r\ \mbox{model},\\
      g&D_r\ \mbox{model},
   \end{array}
   \right.
   \quad
   \check{s}_{jk}\equiv\check{s}_{e_j-e_k},
   \quad
   \bar{\check{s}}_{jk}\equiv\check{s}_{e_j+e_k}.
\end{equation}
By repeating them we obtain
\begin{eqnarray}
   [\ell_j^+\ell_j^-,\ell_k^+\ell_k^-]&=&-2\omega\tilde{g}
    [\ell_j^+\ell_j^-,m_{jk}],\quad m_{jk}\equiv
   \check{s}_{jk}+\bar{\check{s}}_{jk}=m_{kj}, \quad j\ne
   k,\label{elpmcom}\\
   \ [(\ell_j^+\ell_j^-)^n,\ell_k^+\ell_k^-]&=&-2\omega\tilde{g}
   [(\ell_j^+\ell_j^-)^n,m_{jk}]=
   +2\omega\tilde{g}
   [(\ell_k^+\ell_k^-)^n,m_{jk}],\label{n1}\\
   \ [\ell_j^+\ell_j^-,(\ell_k^+\ell_k^-)^m]&=&+2\omega\tilde{g}
   [(\ell_k^+\ell_k^-)^m,m_{jk}]=
   -2\omega\tilde{g}
   [(\ell_j^+\ell_j^-)^m,m_{jk}].
   \label{1m}
\end{eqnarray}
Here and later the identity
\([(\ell_k^+\ell_k^-)^t,m_{jk}]=-[(\ell_j^+\ell_j^-)^t,m_{jk}]\)
is used repeatedly.
Then (\ref{n1}) leads to
\begin{equation}
   \ [(\ell_j^+\ell_j^-)^n,(\ell_k^+\ell_k^-)^m]=+2\omega\tilde{g}
   \sum_{t=0}^{m-1}\left((\ell_k^+\ell_k^-)^{t+n}m_{jk}
   (\ell_k^+\ell_k^-)^{m-t-1}
   -(\ell_k^+\ell_k^-)^{t}m_{jk}
   (\ell_k^+\ell_k^-)^{m+n-t-1}\right)
   \label{nfirst}
\end{equation}
and (\ref{1m}) to
\begin{equation}
   \ [(\ell_j^+\ell_j^-)^n,(\ell_k^+\ell_k^-)^m]=-2\omega\tilde{g}
   \sum_{t=0}^{n-1}\left((\ell_j^+\ell_j^-)^{t+m}m_{jk}
   (\ell_j^+\ell_j^-)^{n-t-1}
   -(\ell_j^+\ell_j^-)^{t}m_{jk}
   (\ell_j^+\ell_j^-)^{m+n-t-1}\right).
   \label{mfirst}
\end{equation}
Summing over \(j\) and \(k\) and adding (\ref{nfirst}) and
(\ref{mfirst})  together with the interchange of the
dummy indices
\(j\leftrightarrow k\) in the latter produces
\begin{equation}
   2[{}^{\ell}Q_n,{}^{\ell}Q_m]=
   +2\omega\tilde{g}
   \sum_{t=0}^{n+m-1}\sum_{j,k}\left((\ell_k^+\ell_k^-)^{t}m_{jk}
   (\ell_k^+\ell_k^-)^{n+m-t-1}
   -(\ell_k^+\ell_k^-)^{t}m_{jk}
   (\ell_k^+\ell_k^-)^{n+m-t-1}\right)=0.
\end{equation}
Thus we obtain
\begin{equation}
   [Q_n,Q_m]=0,\quad n,m=1,\ldots,r
\end{equation}
on the Fock space of Coxeter invariant states.

\bigskip
\subsection{Lax pair with spectral parameter}
\label{laxxi}
In the theory of classical Calogero-Moser models, Lax pair
with spectral parameter (\(\xi\)) plays an important role,
in particular, in elliptic potential models
\cite{Krichever,DHoker_Phong,bcs1,bcs2} for derivation of
spectral curves, etc.  In quantum theory, however, the meaning
and use of the spectral parameter are yet to be established,
partly because of the underdeveloped stage of the quantum models
with elliptic potentials.
Here we   point out a small use of the quantum Lax pair with spectral
parameter in the quantum model with trigonometric potential.
Namely, it accounts for
the useful trick by Polychronakos
\cite{Pol} for the proof of the involution of conserved
quantities in trigonometric
\(A_r\) model.
(Now we have a universal proof of involution for type I, II and III
models, see the previous subsection.)

From the theory of the generalised Lax pair \cite{bcs2} and its
quantum version for degenerate potential models \cite{bms}, we
find that the \(L\) operator can contain one additional complex
parameter \(\xi\):
\begin{equation}
   L^{\xi} = p\cdot\hat{H}+X^{\xi},\quad
   X^{\xi}=i\sum_{\rho\in\Delta_{+}}g_{|\rho|}
   \,(\rho\cdot\hat{H})\left(x(\rho\cdot
   q)-x(\rho^\vee\!\cdot\hat{H}\xi)\right)\hat{s}_{\rho},
   \label{LaxOpDefxi}
\end{equation}
in which the function \(x\) are given in Table I for the degenerate
potentials. With the same \(M\) operator as before, the quantum
equations of motion can be written in a matrix form:
\begin{equation}
   \label{LaxEquationxi}
   {d\over dt}{L^\xi}=i[{\cal H},L^\xi]=[L^\xi,{M}].
\end{equation}
In other words, the \(\xi\) dependent part decouples.
This allows us to define a one parameter family of conserved
quantities
\begin{equation}
   {}^{\xi}Q_n=\mbox{Ts}(L^{\xi})^n,\quad n=1,2,\ldots,
\end{equation}
which turns out to be a \(\xi\) dependent sum of \(Q_n\) and the
lower order conserved quantities \(Q_m\), \(m< n\).
A special limit \(\xi\to-i\infty\) in the trigonometric models
provides a convenient combination which allows easy proof of
involution in the \(A_r\) model.
(In the rational model, the limit reduces to the Lax pair without
spectral parameter. In the hyperbolic models this
limit is ill-defined.)
In the rest of this subsection we consider only the trigonometric
potential models.
Let us denote the limiting \(L^\xi\) operator by \(L^\infty\) which
reads
\begin{equation}
   L^\infty=L+a\sum_{\rho\in\Delta_+}g_{|\rho|}|\rho\cdot\hat{H}|
   \hat{s}_{\rho}
\end{equation}
and the corresponding \(\ell\) operators are given by
\begin{equation}
   \ell^\infty_\mu=\ell_\mu +t_\mu,\quad
   t_\mu=a\sum_{\rho\in\Delta_+}g_{|\rho|}|\rho\cdot\mu|
   \check{s}_{\rho}.
\end{equation}

For the \(A_r\) model in the vector representation
\({\cal R}=\{\mu_j\in{\bf R}^r,
j=1,\ldots,r+1\}\)
with the standard normalisation of roots \(\rho^2=2\), the above expression
simplifies to
\begin{equation}
   \ell^\infty_j=\ell_j +t_j,\quad
   t_j=ag\sum_{k\neq j}
   \check{s}_{jk},
\end{equation}
in which as before we abbreviate \(\ell_{\mu_j}\) as \(\ell_{j}\)
and \(\check{s}_{jk}\equiv\check{s}_{\mu_j-\mu_k}\).
These are the operators introduced in \cite{Pol},
\(\ell_j\leftrightarrow \pi_j\),
\(\ell_j^\infty\leftrightarrow \tilde\pi_j\).
The Lax operator with the spectral parameter gives an `explanation' for the
rather ad hoc introduction of \(\tilde\pi_j\). It is straightforward to show
\begin{equation}
   [\ell_j,\ell_k]=[t_j,t_k],
\end{equation}
which leads to
\begin{equation}
   [\ell^\infty_j,\ell^\infty_k]=[\ell^\infty_j,t_k]+[t_j,\ell^\infty_k]
   =2ag[\ell^\infty_j,\check{s}_{jk}]
   =-2ag[\ell^\infty_k,\check{s}_{jk}].
\end{equation}
This has the same structure as (\ref{elpmcom}) in the previous subsection.
By repeating the same argument we arrive at \cite{Pol}
\begin{equation}
   [{}^\infty\! Q_n,{}^\infty\! Q_m]=0,\quad n,m=1,\ldots,r,\quad
   {}^\infty\! Q_n=\sum_{j=1}^r(\ell^\infty_j)^n,
\end{equation}
which then imply the involution of the conserved quantities obtained from
the original \(L\) operator
\begin{equation}
   [Q_n,Q_m]=0,\quad n,m=1,\ldots,r.
\end{equation}

%%%%%%%%%%%%%%%%%%%%%%%%%%

\section{Summary and comments}
\label{comdis}
\setcounter{equation}{0}
We have discussed various issues related to quantum integrability of
Calogero-Moser models based on all root systems.
These are construction of quantum conserved quantities and a unified proof
of their involution, the relationship between the Lax pair  and
the differential-reflection (Dunkl) operators formalisms, construction of
excited states by creation operators, etc.
They are mainly generalisations of the results known for the models
based on \(A_r\) root systems, which are shown to apply to the
models based on any root systems.
There are some interesting works discussing the integrability issues
of the models based on other classical root systems and the exceptional
ones including the non-crystallographic models
\cite{G2sep}, \cite{G2new}-\cite{Kha}.

Here we list some comments on interesting issues which are not treated in
this paper.
The structure and properties of the eigenfunctions of the trigonometric
potential models, which are generalisations of the Jack polynomials
\cite{Stan}-\cite{Awat}, will be discussed in future publications.
A comprehensive treatment of Liouville integrability of
rational models with harmonic force is wanted.
Our starting point, the factorised Hamiltonian (\ref{facMHamiltonian})
for degenerate potential models, is closely related with supersymmetry
and shape invariance \cite{Eft,Kha}. Further investigation in this
direction is a future problem.
It is a great challenge to formulate various aspects of
quantum Calogero-Moser models with elliptic potentials;
Lax pair, the differential-reflection operators \cite{Cher,Oshima},
conserved quantities, supersymmetry and excited states wavefunctions.

\section*{Acknowledgements}
\setcounter{equation}{0}
We thank K.\, Takasaki, S.\, Odake and P.\, Ghosh for useful discussion.
This work is partially supported  by the Grant-in-aid from
the   Ministry of Education, Science and Culture,
priority area (\#707)  ``Supersymmetry and unified theory of elementary
particles". S.\,P.\,K. and A.\,J.\,P. are supported by the
Japan Society for the
Promotion of Science.
%%%%%%%%%%%%%%%%%%%%%%%
%\appendix
\section*{Appendix A: Root Systems}
\setcounter{equation}{0}

\renewcommand{\theequation}{A.\arabic{equation}}

In this Appendix we recapitulate the rudimentary facts of the root systems
and
reflections to be used in the main text.
The set of roots \(\Delta\) is invariant under reflections
in the hyperplane perpendicular to each
vector in \(\Delta\).  In other words,
\begin{equation}
   s_{\alpha}(\beta)\in\Delta,\quad\forall \alpha,\beta\in\Delta,
\end{equation}
where
%\begin{equation}
%   s_{\alpha}(\beta)=\beta-2(\alpha\cdot\beta/|\alpha|^{2})\alpha.
%\end{equation}
%Dual roots are defined by \(\alpha^{\vee}=2\alpha/|\alpha|^{2}\), in
%terms of which
\begin{equation}
   \label{Root_reflection}
   s_{\alpha}(\beta)=\beta-(\alpha^{\vee}\!\!\cdot\beta)\alpha,
\quad \alpha^{\vee}\equiv2\alpha/|\alpha|^{2}.
\end{equation}
The set of reflections \(\{s_{\alpha},\,\alpha\in\Delta\}\) generates a
group \(G_{\Delta}\), known as a Coxeter group, or finite reflection group.
The orbit of \(\beta\in\Delta\) is the set of root vectors
resulting from the action of the Coxeter group on it.
The set of positive roots \(\Delta_{+}\) may be defined in terms of a
vector \(U\in\mathbf{R}^{r}\), with
\(\alpha\cdot U \neq 0,\,\forall\alpha\in\Delta\), as
those roots \(\alpha\in\Delta\) such that \(\alpha\cdot U>0\).  Given
\(\Delta_{+}\), there is a unique
set of \(r\) simple roots \(\Pi = \{\alpha_{j},\,j=1,\ldots, r\}\)
defined such that they span
the root space and the coefficients \(\{a_{j}\}\) in
\(\beta=\sum_{j=1}^{r}a_{j}\alpha_{j}\) for \(\beta\in\Delta_{+}\)
are all non-negative. The highest root  \(\alpha_h\), for which
\(\sum_{j=1}^{r}a_{j}\) is maximal, is then also determined
uniquely.
The subset of reflections \(\{s_{\alpha},\,\alpha\in\Pi\}\)
in fact generates the
Coxeter group \(G_{\Delta}\).  The products of
\(s_{\alpha}\), with \(\alpha\in \Pi\), are subject solely to the relations
\begin{equation}
   \label{Coxeter_relations}
   (s_{\alpha}s_{\beta})^{m(\alpha,\beta)}=1,\qquad \alpha,\beta\in \Pi.
\end{equation}
The interpretation is that \(s_{\alpha}s_{\beta}\) is a rotation in some
plane by \(2\pi/{m(\alpha,\beta)}\).
The set of positive integers \(m(\alpha,\beta)\)
(with \(m(\alpha,\alpha)=1,\,\forall \alpha\in \Pi\))
uniquely specify the Coxeter group.
The weight lattice \(\Lambda(\Delta)\) is defined as the \(\mathbf{Z}\)-span
of the fundamental weights \(\{\lambda_j\}\), \(j=1,\ldots,r\), defined
by
\begin{equation}
   \alpha^{\vee}_j\!\!\cdot\lambda_k=\delta_{jk},\quad \alpha_j\in\Pi.
\end{equation}

The root systems for finite reflection groups may be divided into two
types: crystallographic and non-crystallographic.
Crystallographic root systems satisfy the additional condition
\begin{equation}
   \alpha^{\vee}\!\!\cdot\beta\in\mathbf{Z},\quad \forall
   \alpha,\beta\in\Delta,
\end{equation}
which implies that the \(\mathbf{Z}\)-span of \(\Pi\) is a lattice in
\(\mathbf{R}^{r}\) and  contains all roots in \(\Delta\).
We call this the root lattice, which is denoted by \(L(\Delta)\).
These root systems are associated with simple Lie
algebras: \{\(A_{r},\,r\ge 1\}\), \(\{B_{r},\,r\ge 2\}\), \(\{C_{r},\,r\ge
2\}\),
\(\{D_{r},\,r\ge 4\}\), \(E_{6}\), \(E_{7}\), \(E_{8}\), \(F_{4}\) and
\(G_{2}\).  The Coxeter groups for these root
systems are called Weyl groups.  The remaining non-crystallographic root
systems %\cite{Coxeter_groups}
are \(H_{3}\), \(H_{4}\), whose Coxeter groups are the
symmetry groups of the icosahedron and four-dimensional 600-cell,
respectively,
and the dihedral group of order \(2m\), \(\{I_{2}(m),\,m\ge 4\}\).

\section*{Appendix B: Conserved quantities}
\setcounter{equation}{0}
\label{consquant}
\renewcommand{\theequation}{B.\arabic{equation}}
Here we list for each root system how the full set of independent
conserved quantities are obtained by choosing proper representations of the
Lax pair.
We choose those of the lowest dimensionality for the convenience of
practical calculation.
Of course there are many other choices of representations giving equally
good sets of conserved quantities.
The independence of the conserved quantities can be easily verified by
considering the free limit: \(g_{|\rho|}\to0\).
\begin{enumerate}
\item
\(A_r\): For all powers, the vector representation (\(r+1\) dimensions) is
enough.
\item
\(B_r\): For all powers, the  representation consisting of short roots
\(\{\pm e_j: j=1,\ldots,r\}\), (\(2r\) dimensions) is enough.
Here we adopt the following explicit
parametrisation of the \(B_r\) root system:
\begin{equation}
   B_r\ \mbox{root system :}\quad \Delta=\{\pm e_j\pm e_k,\
    \pm e_j, \quad j,k=1,\ldots,r|
   e_j\in{\bf R}^r, e_j\cdot e_k=\delta_{jk}\}.
   \label{brroots}
\end{equation}
\item
\(C_r\): For all powers, the  representation consisting of long roots
\(\{\pm 2e_j: j=1,\ldots,r\}\), (\(2r\) dimensions) is enough.
The following
parametrisation of the root system is used:
\begin{equation}
   C_r\ \mbox{root system :}\quad \Delta=\{\pm e_j\pm e_k,\
    \pm 2e_j, \quad j,k=1,\ldots,r|
   e_j\in{\bf R}^r, e_j\cdot e_k=\delta_{jk}\}.
\end{equation}
\item
\(D_r\): For all even powers, the vector representation (\(2r\) dimensions)
is enough. For the
additional one occurring at power
\(r\), the (anti)-spinor representation (\(2^{r-1}\) dimensions) would be
necessary. They are minimal representations.
\item
\(E_6\):
For all powers, the {\bf 27} (or \({\bf
\overline{27}}\)) dimensional representation of the Lie algebra
is enough. They are minimal representations.
\item
\(E_7\):
For all powers, the {\bf 56}  dimensional representation of the Lie algebra
is enough. This is a minimal representation.
\item
\(E_8\):
For all powers, the 240  dimensional representation consisting
of all the roots is enough.  This is not the same as the adjoint
representation of the Lie algebra.
\item
\(F_4\):
For all powers, either of the 24  dimensional representation consisting
of all the long roots or the short roots is enough. These are not Lie
algebra representations.
\item
\(G_2\):
For all powers, either of the 6  dimensional representations consisting
of all the long roots or the short roots is enough. These are not Lie
algebra representations.
\item
\(I_2(m)\): For both powers 2 and \(m\),
the representation consisting of \(V_m\)
is enough. Here, \(V_m\) is the set of vectors with `half" angles
of the roots (see (\ref{dihedroots}))
given by
\begin{equation}
   V_m=\{v_j=(\cos((2j-1)\pi/2m),\sin((2j-1)\pi/2m))\in{\bf R}^2|\
   j=1,\ldots, m\}.
   \label{vmpara}
\end{equation}
\item
\(H_3\): For all powers, the representation consisting of all the 30 roots
is enough.
\item
\(H_4\): For all powers, the representation consisting of all the 120 roots
is enough.
\end{enumerate}

%%%%%%%%%%%%%%%%%%%%%%%%%%%%%%%%%%%%%%%%%%%%%%%%%%%%%

\end{document}